\documentclass[twocolumn,showpacs,preprintnumbers,amsmath,amssymb]{revtex4-1}

\usepackage[dvips]{graphicx}
\usepackage{color}
\usepackage{enumerate}

\newcommand{\beq}{\begin{equation}}
\newcommand{\eeq}{\end{equation}}
\newcommand{\bma}{\begin{math}}
\newcommand{\ema}{\end{math}}
\newcommand{\beqa}{\begin{eqnarray}}
\newcommand{\eeqa}{\end{eqnarray}}

\def\mf#1{\mathfrak{#1}}

\def\opone{\le\textbf{}\textbf{}avevmode\hbox{\small1\kern-3.8pt\normalsize1}}

\newcommand{\be}[1]{     \begin{eqnarray} \mbox{$\label{#1}$}   }

\newcommand{\ee}{\end{eqnarray}}
\newcommand{\pref}[1]{(\ref{#1})}

\newcommand\cf {{\it cf.  }}

\begin{document}

\title{Exclusion statistics for quantum Hall states in Tao-Thouless limit}

\author{M. Kardell}

\author{A. Karlhede}

\affiliation{Department of Physics,
Stockholm University \\
AlbaNova University Center\\
SE-106 91 Stockholm, Sweden}

\date{\today}

\begin{abstract}

We consider spin-polarized abelian quantum Hall states in the Tao-Thouless limit, {\it ie} on a thin torus.
For any filling factor $\nu=p/q$ a well-defined sector of low-energy states is identified and the exclusion statistics of the excitations is determined. 
We study numerically, at and near  $\nu=1/3$ and $2/5$, how the low energy states develop as one moves away from the TT-limit towards the physical regime. 
We find that the lowest energy states in the physical regime develop from states in the low energy sector but that the exclusion statistics is modified.  
\end{abstract}

\maketitle

\section{Introduction}\label{Intro}

Polarized interacting electrons in the lowest Landau level  in the Tao-Thouless limit, which corresponds to a thin torus or cylinder, has been a subject of recent study, see {\it eg} Ref. \onlinecite{bk1,Lee05,bk2, emil,Jansen}. 
In this limit the problem becomes equivalent to a one-dimensional lattice model with a purely electrostatic repulsive interaction between the electrons. The hamiltonian can be 
diagonalized for any filling factor $\nu=p/q$ and the energy eigenstates are the states where the electrons have fixed positions on the lattice\cite{bk2}. 
In the ground state, the Tao-Thouless state, the electrons are as far separated as possible; the state has a unit cell with $p$ electrons on $q$ sites and is $q$-fold degenerate. 
There is a gap to all excitations and the low energy charged excitations are domain walls between the degenerate ground states; these domain walls have charge $\pm e/q$. 
The Tao-Thouless states, and their excitations, are limits of  hierarchy wave functions describing abelian quantum Hall states and there
is strong evidence that they develop adiabatically into such states as the circumference of the torus increases in cases where such states are observed. The hierarchy 
construction of quantum Hall states is manifest in the TT-limit: each state is obtained by condensing quasiparticles, {\it ie} domain walls, in a parent state\cite{wehierarchy}. In our opinion, the TT-states
are abelian quantum Hall states. 

In this article we consider the low-energy excitations in the TT-limit in general.
It was previously noted that the unit cell $C$ for the TT  qround state can be written as $C=\mf {hp}$ where $\mf p$  is the quasielectron  and $\mf h$ is the quasihole excitation \cite{emil}.
Thus the ground state is a string $\mf {hphphphphphp\dots}$. Interchanging a nearest neighbor particle and hole gives $\mf {hphphhpphphp\dots}$.
This state has two nearby domain walls, $\mf{hh}$ and $\mf{pp}$---these are the quasiparticles with charges $\pm e/q$. 
Here we show that, given a certain condition on the electron-electron interaction, the low-energy sector of states consists of all reorderings of the $\mf p$'s and $\mf h$'s that make up the ground state. 

We study how the low-energy states develop away from the TT-limit using exact diagonalisation at and around $\nu=1/3$ and $2/5$. 
We find that the low energy sector for any  circumference consists of states that develop from the low energy states in the TT-limit. 

The exclusion statistics \cite{exclusion} was determined for $\nu=1/3$ in the TT-limit in Ref. \onlinecite{we}. 
We here generalize this to arbitrary filling factor $\nu=p/q$ and find a simple result for the exclusion statistics,
which, however, in general differs from the values established for the excitations in the physical regime\cite{canright,canright94,wu94,su,isakov,sitko03}.
In particular it disagrees in certain cases with the values obtained by assuming the quasiparticles are anyons and counting anyon wave functions. 
For example, at filling factors 1/3 and 2/5 the diagonal statistics in the TT-limit agrees with that in the physical regime for the holes but differs for the particles.
We study numerically for these filling factors how the exclusion statistics for the quasiparticles change as the circumference increases and the values that are consistent with the particles being anyons are recovered. 
A recent discussion of exclusion statistics in connection to entanglement spectra can be found in Ref. \onlinecite{hermanns10}, see also Ref. \onlinecite{haldane08,bergholtz10}.

The quasiparticles in the TT-limit are domain walls and as a consequence the exchange statistics cannot be calculated in this limit\cite{we}. However, the exchange statistics can be considered
by making use of the TT-limit and invoking the adiabatic continuity to the bulk quasiparticles\cite{SeidelLee,Seidel08}.

In Sec. \ref{Low_section} we determine the low-energy states in the TT-limit and in Sec. \ref{Numerics} we investigate to what extent these results hold away from the TT-limit by numerically investigating small systems.
The exclusion statistics is discussed in Sec. \ref{exclusion} and our findings are discussed in Sec. \ref{discussion}.
Derivations and details of our analysis are given in three appendices. 

\section{Low-energy states in TT-limit}\label{Low_section}

In this section we identify the space of low-energy states in the Tao-Thouless limit for arbitrary filling factor $\nu=p/q\le 1$. Mathematical details are given in the 
appendices and details about the TT-limit can be found in  Ref. \onlinecite{emil}. 

We consider a single Landau level of spin polarized electrons on a torus with lengths $L_1,L_2$.
Using Landau gauge, the natural one-particle eigenstates $\psi_k$, $k=1,\dots N_s$,
are gaussians concentrated along circles around the torus (in the $L_2$-direction say, $k$ is the momentum in this direction).
This maps the Landau level onto a one-dimensional lattice model with $N_s=L_1L_2/2\pi$ sites numbered by $k$,
where each site is either empty or occupied by an electron.
The lattice constant is $2\pi/L_1$ in units of the magnetic length $\ell=\sqrt{\hbar c/eB}$.
A basis of many-particle states is given by $|n_1 n_2 \dots n_{N_s}\rangle$, where $n_i=0,1$,
and $N=\sum_i^{N_s}n_i$ is the number of electrons;
the filling factor is  $\nu=p/q=N/N_s$.
Since the position $k$ gives the conserved momentum, it follows that a general translationally invariant 
two-body hamiltonian consists of all terms where two electrons hop while preserving the position of their center of mass.
Let $V_{km}$ be the amplitude for this process when the separations between the electrons are $k\pm m$.
When $L_1 \rightarrow 0$, the lattice constant goes to infinity and the overlap between the single-particle states with different $k$ vanishes.
As a consequence, only the electrostatic terms $V_{k0}$ remain.
This is the thin torus, or Tao-Thouless, limit.
It should be noted that one can think of this limit as simply being a change of the hamiltonian, {\it ie}  the $V_{km}$, while keeping the configuration space unchanged.
Thus the TT-limit does describe a two-dimensional QH system, albeit with a peculiar anisotropic interaction.

In the TT-limit the interacting electron problem can be solved exactly since the interaction is purely electrostatic:
The energy eigenstates are simply the states $|n_1 n_2 \dots n_{N_s}\rangle$.
For an interaction that obeys the concavity condition $V''_k\equiv V_{k-1,0}+V_{k+1,0}-2V_{k0}>0$,
the ground state and the low energy excitations have the same qualitative properties as the standard abelian QH states.
The $q$-fold degenerate ground state is  crystal-like with a unit cell of length $q$ containing $p$ electrons.
There is a gap to excitations and the lowest energy charged excitations have charge $\pm e/q$ and are the domain walls between the degenerate ground states. 

The unit cell $C$ for the ground state is obtained by considering a circle with $q$ equidistant sites and $p$ equidistant electrons.
Letting each electron relax to the closest site gives the unit cell\cite{emil}. 
Electron $j$ is then placed at site $x_j$, where 
\begin{equation}\label{relaxation}
x_{j}\equiv_q\left[\frac{j}{p}q+\delta\right] \ . 
\end{equation}
Here, $\equiv_q$ denotes equality modulo $q$ and $\left[\dots\right]$ is the rounding function: 
$\left[n+\epsilon\right]=n$ if $-\frac{1}{2}\leq\epsilon<\frac{1}{2}$; the  parameter $\delta$ corresponds to the freedom to rotate 
all the initial (unrelaxed)  electron positions by $\delta$, or, equivalently, to rotate the sites by $-\delta$.

There is strong evidence that the TT-ground state,
given by \eqref{relaxation} at  $\nu=p/q$, $q$ odd,
is adiabatically connected to the standard ground state that describes an abelian QH state for a realistic interaction.
We claim that the TT-states \pref{relaxation} are abelian qauntum Hall states.
The hierarchy construction of QH states is proven  in the TT-limit:
Each TT-ground state is a condensate of quasiparticles in another parent TT-ground state.
The unit cell ${ C}$ can be represented in $q$ different ways by translating the cell (imposing periodic boundary conditions).
Transposing the cell gives a cell that is a translation of the original one.
It is convenient to introduce a standard representation ${ C}_R$ (${ C}_L$) which is the cell where the electrons are as far to the right (left) as possibile,
{\it ie} the cell with the maximal (minimal) value of $\sum_j x_j$. 

The ground state unit cell at $\nu=p/q$ can be written as $C_{R}=\mf{hp}$, see \ref{app:ph-properties}, where $\mf{h}$ 
and  $\mf{p}$ are the quasihole and quasielectron with charges $\pm e/q$. When inserting $\mf{p}$ or $\mf{h}$  
between two unit cells in the ground state they produce domain walls with charges $\pm e /q$. 
The number of sites, $l_{\pm}$, and the number of electrons, $n_{\pm}$,  in the quasiparticles are given by
\begin{eqnarray}\label{ln}
l_{\pm}&\equiv_{q}&\mp p^{\varphi\left(q\right)-1} \  \nonumber \\
n_{\pm}&\equiv_{p}&\pm q^{\varphi\left(p\right)-1} \ ,
\end{eqnarray}
where $0<l_{\pm},n_{\pm}\le q $; throughout the article we use $+$ and $\mf p$ interchangeably to denote the quasielectrons (and $-$ and $\mf h$ to denote the quasiholes). 
These relations are derived in \ref{app:ph-properties}.
Here, $\varphi(n)$ is Euler's totient function. 
($\varphi(n)$ is the number of positive integers less than or equal to $n$ that are coprime to $n$, {\it ie} have no common positive factor with $n$ other than 1.)
$l_+$ and $l_-$ are the lengths  of ({\it ie} the number of sites in) the quasielectron and quasihole in the TT-limit; $l_++l_-=q$.
For example, for $\nu=1/3$, 2/5, 4/11 and 5/13 one finds $(l_+,l_-)=\left(2,1\right)$, $(2,3)$,
$(8,3)$ and $(5,8)$ respectively. 
Moreover, due to the hierarchical structure, $\mf{h}$ and  $\mf{p}$ are themselves the unit cells at $\nu_\pm=n_\pm/l_\pm$.
This means that they are given by the relaxation procedure above. 

Let us consider the two examples $\nu=2/5$ and $\nu=5/13$. For $\nu=2/5$,
choosing $\delta=0$ in  \pref{relaxation}, we find the positions of the electrons in the unit cell to be:
$x_j=3,5$, corresponding to $C_R=00101$ (since $\sum_j x_j$ is maximal this is in the $R$-representation).
From \pref{ln} we find that the quasielectron and quasihole have lengths $l_+=2$, $l_-=3$ and contain $n_+=n_-=1$ electrons respectively, hence $\mf p=01$ and  $\mf h=001$.
Note that 001 is the unit cell for the ground state at $\nu=1/3$ as well as the quasihole at $\nu=2/5$ (the corresponding result holds for 01 and 1/2,
although the latter does not continue to a QH state since $q=2$ is even).
The cells are written in the $R$-representation and we see that $C_R=\mf h \mf p$.
Inspecting the string $\mf h\mf p=00101$ we see that by moving the first electron one step to the left,
or the last electron one step to the right (assuming periodic boundary conditions),
we obtain a translation of $C_R$.
When these moves are performed in the TT-ground state,
{\it ie} in a string of $C_R=\mf h \mf p$,
one finds that they interchange a nearest neighbor $\mf p$ and $\mf h$;
the first gives $\mf {h p} \rightarrow \mf {p h}$ whereas the second leads to $C_R \rightarrow C_L$.

Repeating the analysis for $\nu =5/13$, we find from \pref{relaxation} (with $\delta=0$) the positions of the electrons in the unit cell to be:
$x_j=3,5,8,10,13$, corresponding to $C=0010100101001$.
Translating $C$ to maximize $\sum_j x_j$ we find the unit cell $C_R=0010010100101$.
From \pref{ln} we find that the quasielectron and quasihole have lengths $l_+=5$, $l_-=8$ and contain $n_+=2$, $n_-=3$ electrons respectively.
Applying the relaxation procedure \pref{relaxation}, then gives $\mf p=00101$ and  $\mf h=00100101$.
Note that 00101 is the unit cell for the ground state at $\nu=2/5$ as well as the quasielectron at $\nu=5/13$ (similarly for 00100101 and 3/8).
We have chosen to write the cells in $R$-representation and we see that $C_R=\mf h \mf p$.
Inspecting the string $\mf h\mf p=0010010100101$ we see that by moving the second electron one step to the left,
or the last electron one step to the right (assuming periodic boundary conditions),
we obtain a translation of $C_R$.
When these moves are performed in the TT-ground state,
one finds, just as for 2/5, that they interchange a nearest neighbor $\mf p$ and $\mf h$;
the first gives $\mf {h p} \rightarrow \mf {p h}$ whereas the second leads to $C_R \rightarrow C_L$.

These results obtained for $\nu =2/5$ and $5/13$ generalize to arbitrary filling factor and will be crucial in identifying the low energy excitations.
We show in \ref{app:ph-properties} that in the unit cell $C_R=\mf h \mf p$ that gives the ground state at $\nu=p/q, \, p>1$
({\it ie} the unit cell that is given by \pref{relaxation})
there are precisely two electrons with the property that moving one of them one step 
(to the left for the one to the left  and to the right for the one to the right)
interchanges a nearest neighbor $\mf p$ and $\mf h$ in the string of $C_R$ that make up the ground state;
moving the left electron gives $\mf {h p} \rightarrow \mf {p h}$ whereas moving the right leads to $\mf {h p} \rightarrow \mf {p h}$.
(The first acts within a unit cell  $C_R=\mf h \mf p\rightarrow \mf p \mf h$, whereas the second acts within a translated unit cell.) 
Furthermore, the energy of these excitations is of the order of $V''_q$; we denote this by ${\cal O}_q$,
and it is the minimum energy excitation of the system (except for the zero energy excitations that translate the whole system giving one of the degenerate ground states). 
All other excitations are of order  ${\cal O}_k \propto V''_k$, $k<q$.
(We assume $V''_k> V''_{k+1}$, in the TT-limit, see below.)
We assume that $\mf p$, $\mf h$ (just as $C_R$) are given in the $R$-representation;
this makes the identification of a certain state given in the $0,1$ notation as a $\mf p, \mf h$ state unique.
Note that interchanging a nearest neighbor $\mf h$ and  $\mf p$ in the ground state,
which is a string of $C_R=\mf{hp}$, corresponds to creating two domain walls,
$\mf {hh}$ and $\mf{pp}$ with charges $\pm e/q$,
next to each other---this is a minimally separated particle-hole pair,
which is a natural minimum energy excitation. For $\nu=1/q$ the above still holds with the exception that there is now one electron per unit cell that can be moved both to the right and 
to the left leading to an interchange of  $\mf p$ and $\mf h$. 

Any interchange $\mf{hp} \leftrightarrow \mf{ph}$,
wherever it takes place in an arbitrary string of $\mf h$ and $\mf p$,
is obtained by a minimum energy hop $01\leftrightarrow 10$ with energy of order ${\cal O}_k, \, k\ge q$ as described above.
This follows since $\mf{hphp}\leftrightarrow\mf{hhpp}$ differ in energy by $V''_q$;
whereas $\mf{hphh}\leftrightarrow\mf{hhph}$ and $\mf{pphp}\leftrightarrow\mf{phpp}$ differ in energy by ${\cal O}_k,\,  k>q$.
The last statement can be understood by noting that the two states are degenerate if the surrounding elements are symmetric around $\mf {ph}$.
The energy difference is brought about by any antisymmetry $k$ sites away.
For a proof of these results, see \ref{app:ph-proofs}.

If $V''_k \gg V''_{k+1}$, for all $k$, then we can identify the low-energy sector as consisting of states that can be described using only $\mf h$ and $\mf p$.
However, several comments are in order.
First, taking the limit of a standard interaction this condition may not be satisfied.
It holds for a screened Coulomb interaction, $V(r)=e^{-\alpha r}/r$ (periodized over the torus) but not for the unscreened case, $V(r)=1/r$, see Fig. \ref{fig:5_17_yuk} below.
(Because of the periodization in the $L_1-$direction, the interaction in the TT-limit,
$L_1\rightarrow 0$,  becomes the electrostatic interaction between two homogeneously charged parallel lines,
giving $V_{k0}\sim \ln k$ for the unscreened Coulomb case and $V_{k0}\sim g(k)e^{-\beta k}$ for the screened case, where $g(k)$ is some non-exponential function.)
What one finds quite generally though is that $V''_k > V''_{k+1}$.
Second, while the specific form of the interaction in the physical regime is well-defined,
this is not the case in the TT-limit---one may simply modify the interaction in this limit so that the condition is fulfilled.
Third, even if $V''_k \gg V''_{k+1}$ it may be that a sufficiently large number of $\mf p$, $\mf h$ excitations has a higher energy than the lowest energy states that are not of this type.
Bearing the provisos above in mind, we identify the low-energy sector as consisting of states that can be described using only $\mf h$ and $\mf p$.
One further argument for this is of course that it is in accord with the expectation in the physical regime. 
As shown in \ref{app:ph-proofs} these states---and no others---differ in energy compared to the ground state by $\mathcal O_q$.

As an illustrative aside we consider the hierarchy construction of 5/13.
Since $5/13=1/(3-1/(2+1/2))$ it follows that 5/13 is obtained by starting from the Laughlin state at 1/3 and condensing quasielectrons to obtain 2/5;
condensing quasiholes in 2/5 then gives 5/13.
In the TT-limit, the 1/3 ground state is 001 and $\mf h=0$, $\mf p =01$.
Inserting $\mf p=01$ in the 001 ground state once per unit cell gives a state with unit cell 00101---this is the 2/5 ground state.
(Note that if one had inserted a lower homogeneous density of $\mf p=01$ one would have obtained the states $(001)_n01$;
these are the $(n+1)/(3n+2)$ ground states.)
One finds from \pref{ln} that the quasiparticle excitations at 2/5 are  $\mf p=01$, $\mf h=001$.
Inserting $\mf h=001$ once per every two unit cells gives $001(00101)_2$---this is the 5/13 ground state.
This illustrates how the hierarchy construction, where states are obtained by successive condensation of quasiparticles, is  manifest in the TT-limit.

\section{Numerics away from TT-limit}\label{Numerics}

In this section we consider how the low-energy sector identified above in the TT-limit develops as $L_1$ increases.
We do this by performing exact diagonalization of small systems at and around filling factors $\nu=1/3,2/5$.
The results are presented in Fig. 1-5.
In each figure the spectra are shown for $0 \le L_1\le \sqrt{2\pi N_s}$ and an unscreened Coulomb interaction, $V(r)=1/r$, that is periodized, in the $L_1$ and $L_2$ directions, over the torus.
The black and red lines are continuations of the low-energy states identified above in the TT-limit,
whereas grey lines correspond to higher energy states in this limit.
Red lines mark the very lowest  states in the TT-limit, where they are identified in terms of particle and/or hole excitations of the ground state.
We use the convention that when two lines with different colors but with the same quantum numbers approach each other and an avoided crossing occurs 
then the lower line is chosen as red or black (in this order of priority) after the crossing.
In each figure $V''_q$ is indicated by a star; this is, as discussed above, a characteristic energy for the low energy excitations in the TT-limit.
Energies are given relative to the energy of the state that has lowest energy in the TT-limit and scaled by $L_1$ to cancel the $1/L_1$ behaviour of $V_{km}$ as $L_1\rightarrow 0$.
The main conclusion from these figures is that for each filling factor the lowest energy states at any $L_1$ develop from states in the low-energy sector in the TT-limit (black or red lines),
without mixing of high-energy TT-states (grey lines). 
There is no obvious simple rule for precisely how many of the lowest states this is true for, but it is always true for some set for all the $\nu=p/q$ that we have studied and we propose that it is true in general.
How strong this restriction on the number of states involved in determining the low energy sector is depends on the filling factor:
At $\nu=1/2$,  since $\mf h=0, \, \mf p=1$ all states are in the low-energy sector in the TT-limit,
whereas the fraction of low-energy states decreases rapidly with increasing $q$. 
 
\begin{figure}
\begin{centering}
\includegraphics[width=0.90\columnwidth]{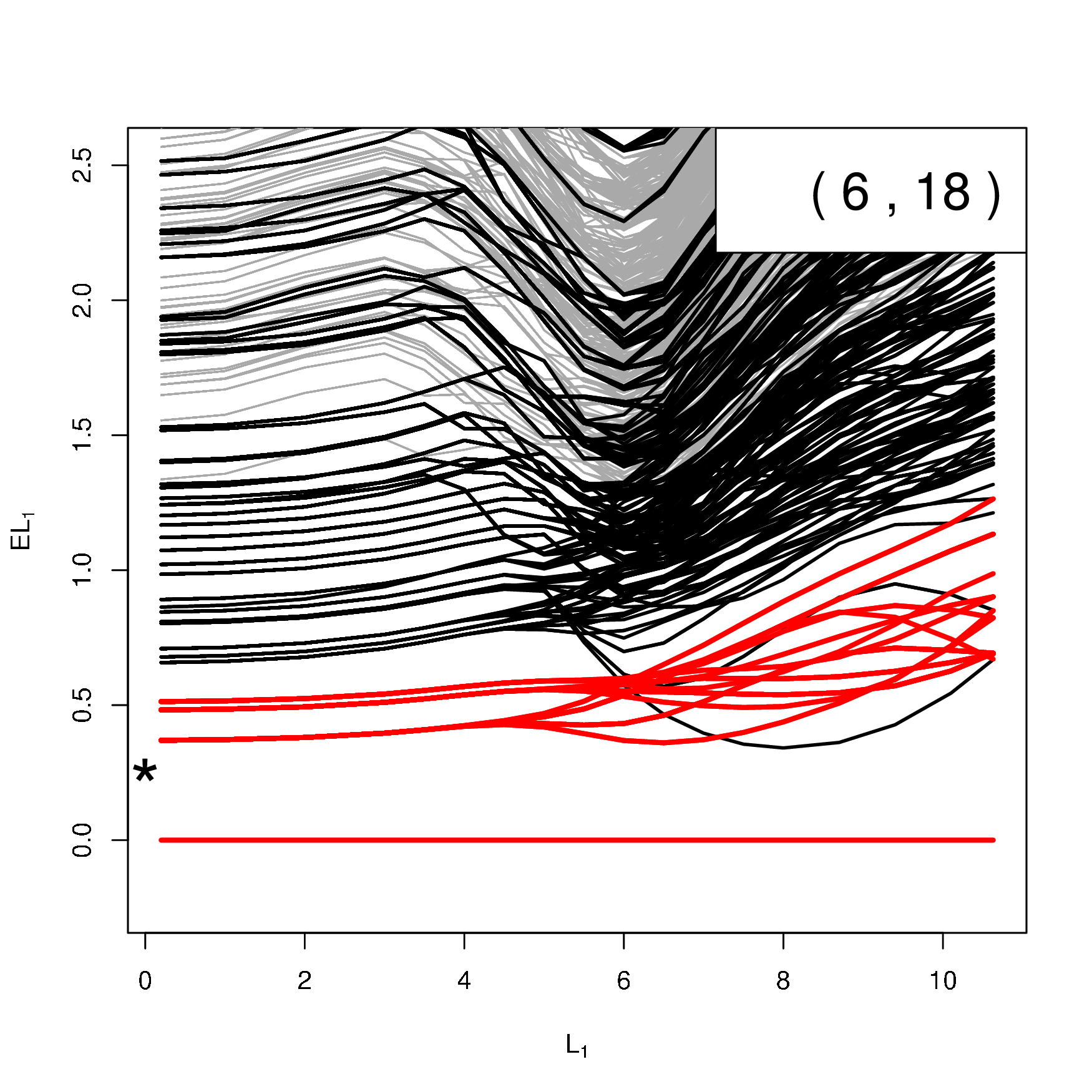}
\includegraphics[width=0.90\columnwidth]{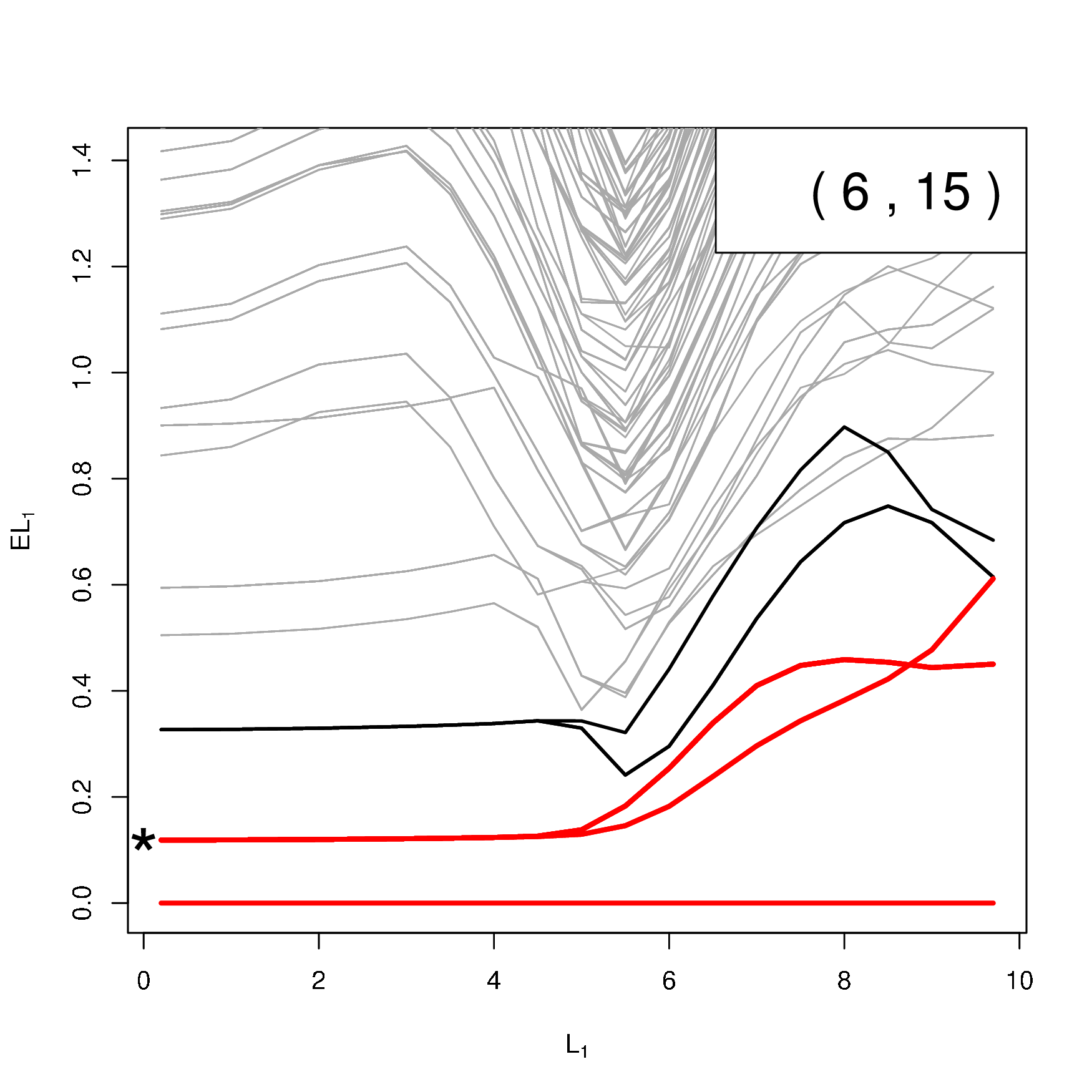}
\par\end{centering}

\caption{Energy spectra for $\nu=1/3$ with six unit cells $(N,N_s)=(6,18)$ (upper panel) and $\nu=2/5$ with three cells $(N,N_s)=(6,15)$ (lower panel).
Energies, scaled by $L_1$, are shown for $0\le L_1\le \sqrt{2\pi N_s}$ and are given relative to the energy of the state that has lowest energy in the TT-limit.  
Red lines denote the ground states and the one particle-hole excitations (identified in the TT-limit).
Black lines denote the remaining low energy states. 
The characteristic variation in the energy of the low-energy excitations in the TT-limit, $V''_q$, is indicated by a star.
(Each state is, at least, 3 and 5-fold degenerate respectively.)
\label{fig:4_10_ph_5_15_ph}}

\end{figure}

Fig. \ref{fig:4_10_ph_5_15_ph} shows the spectra for  $\nu=1/3$ and $2/5$. The ground states and the one particle-hole excitations (identified in the TT-state) are marked in red. 
As noted before \cite{bk2} the ground states develop continuously, without the gap closing.  The one particle-hole excitations remain low energy excitations for all $L_1$, but there are
additional (black) states that mix with these for large $L_1$.  Some of these are presumably states that approach the particle-hole TT-states in the limit $L_1 \rightarrow \infty$ (for fixed $N,N_s$). 

\begin{figure}
\begin{centering}
\includegraphics[width=0.90\columnwidth]{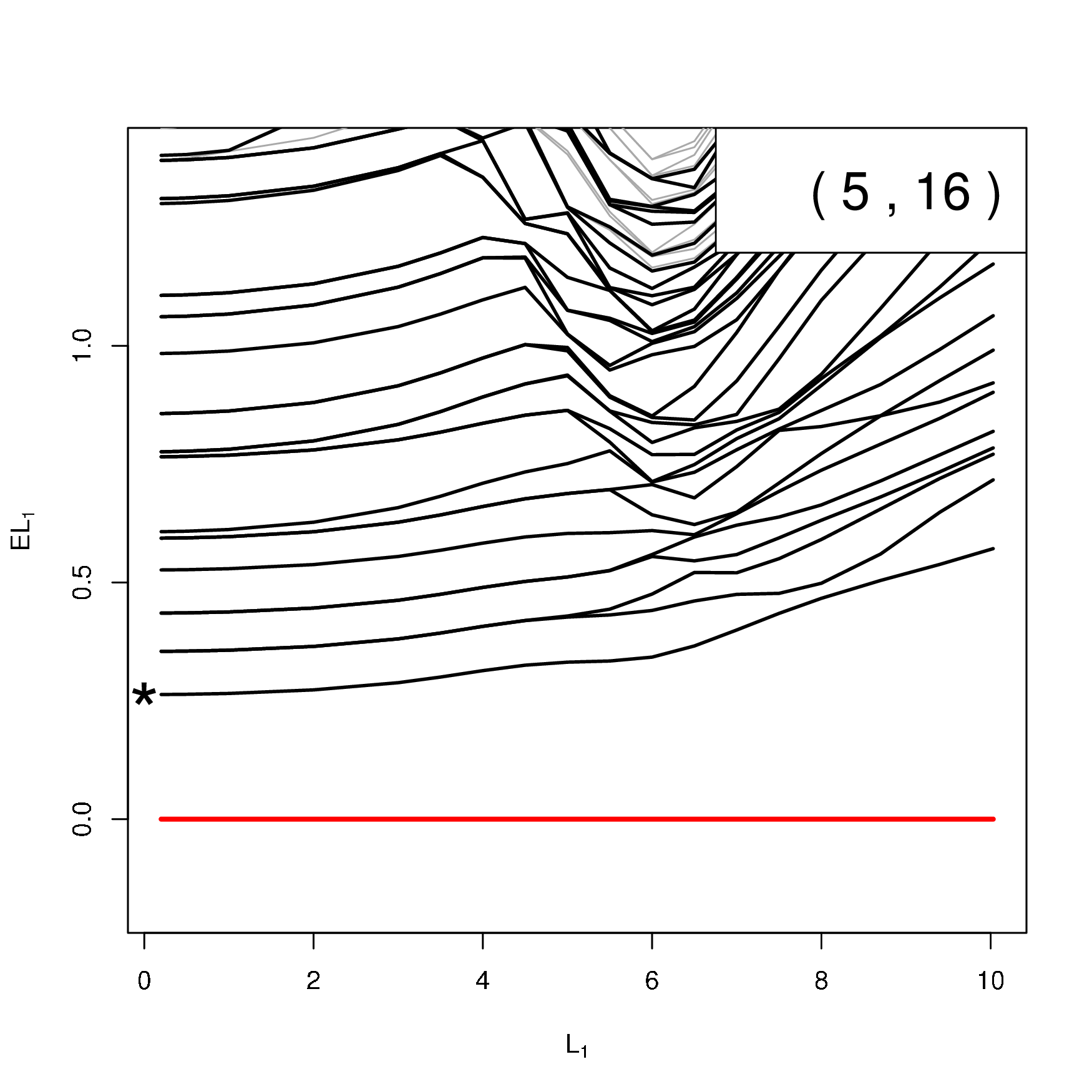}
\includegraphics[width=0.90\columnwidth]{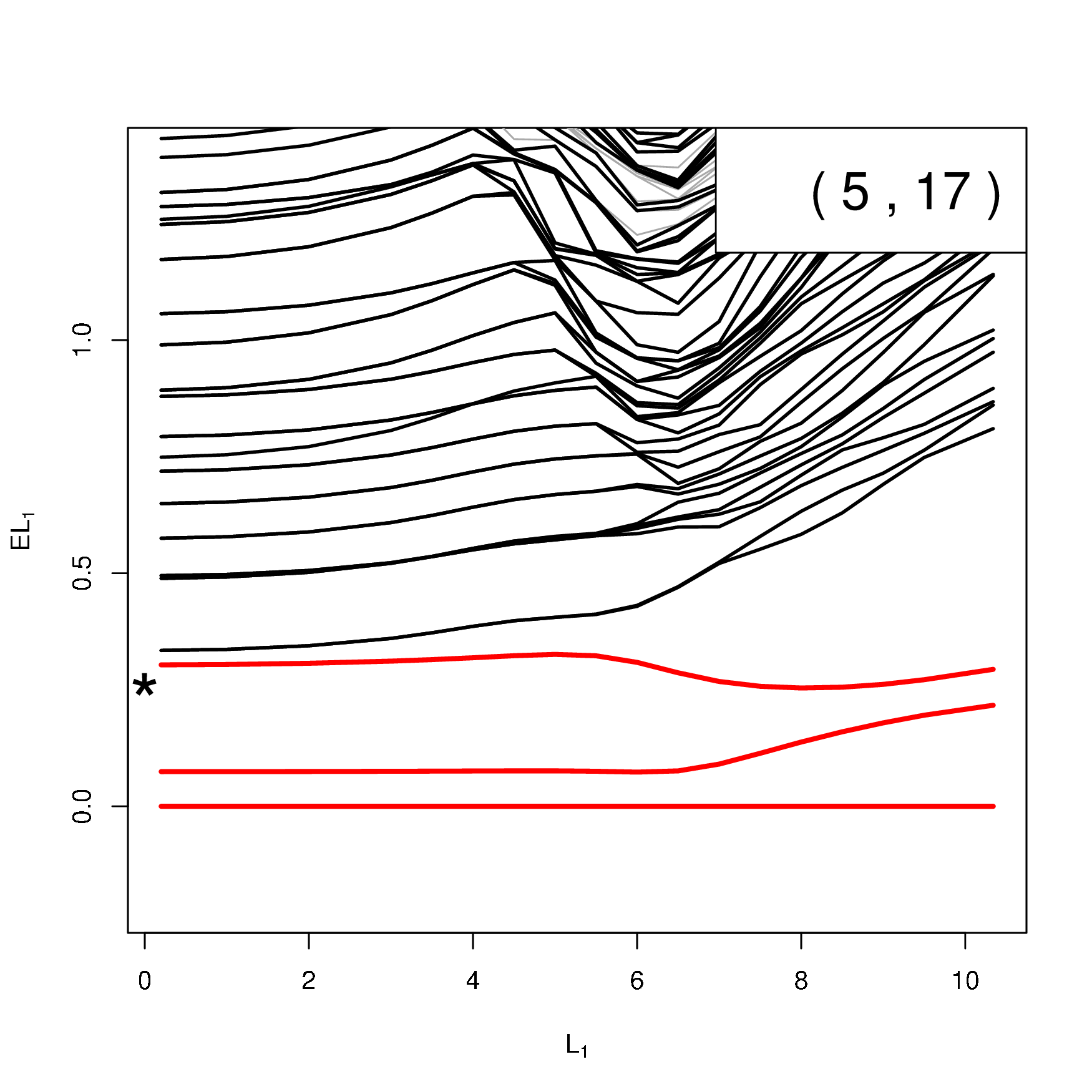}
\par\end{centering}

\caption{
Energy spectra for one $(N,N_s)=(5,16)$ and two (5,17) holes at $\nu=1/3$ with five unit cells---these excitations, identified in the TT-limit, are shown as red lines.
\label{fig:5_15_h_hh}}
\end{figure}

\begin{figure}
\begin{centering}
\includegraphics[width=0.90\columnwidth]{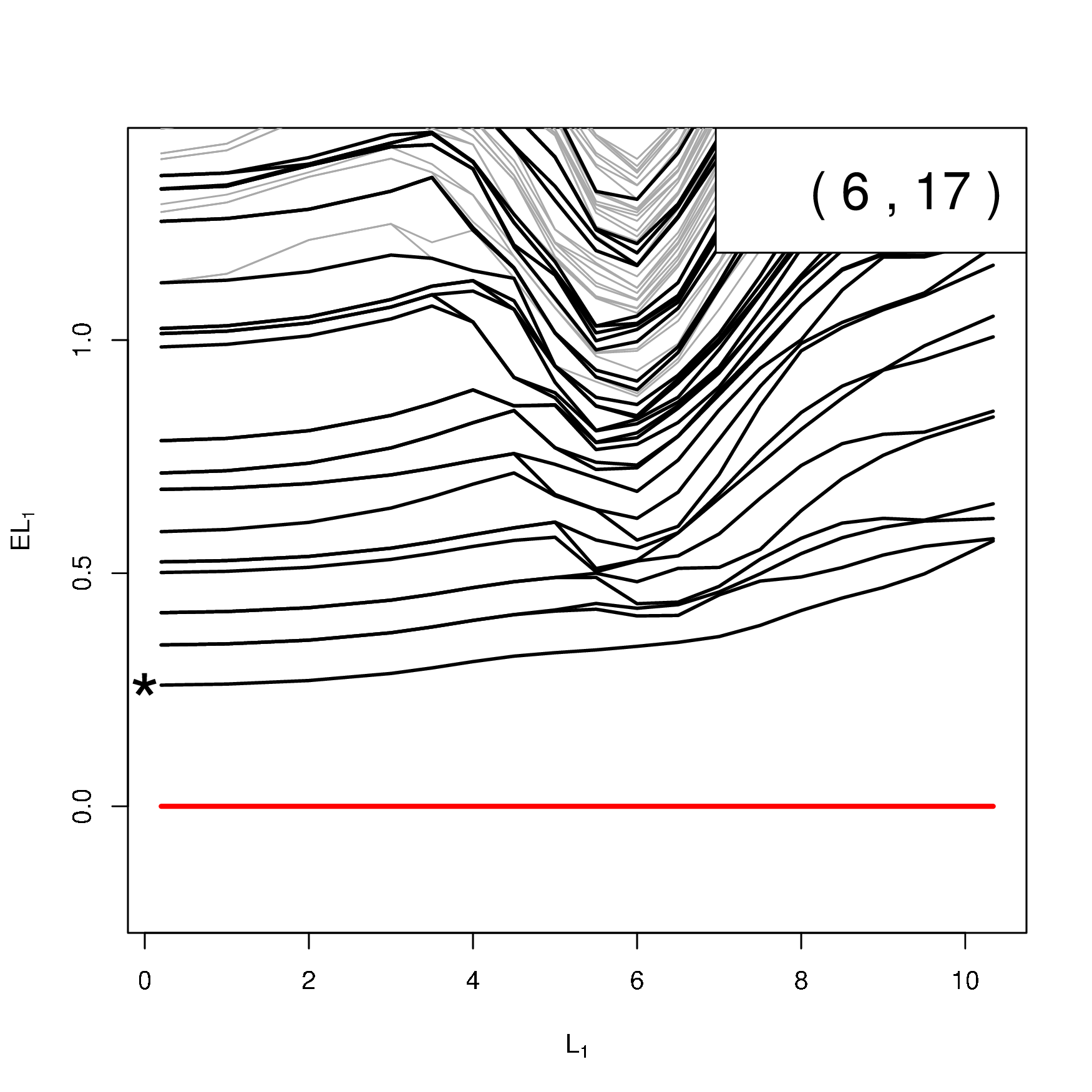}
\includegraphics[width=0.90\columnwidth]{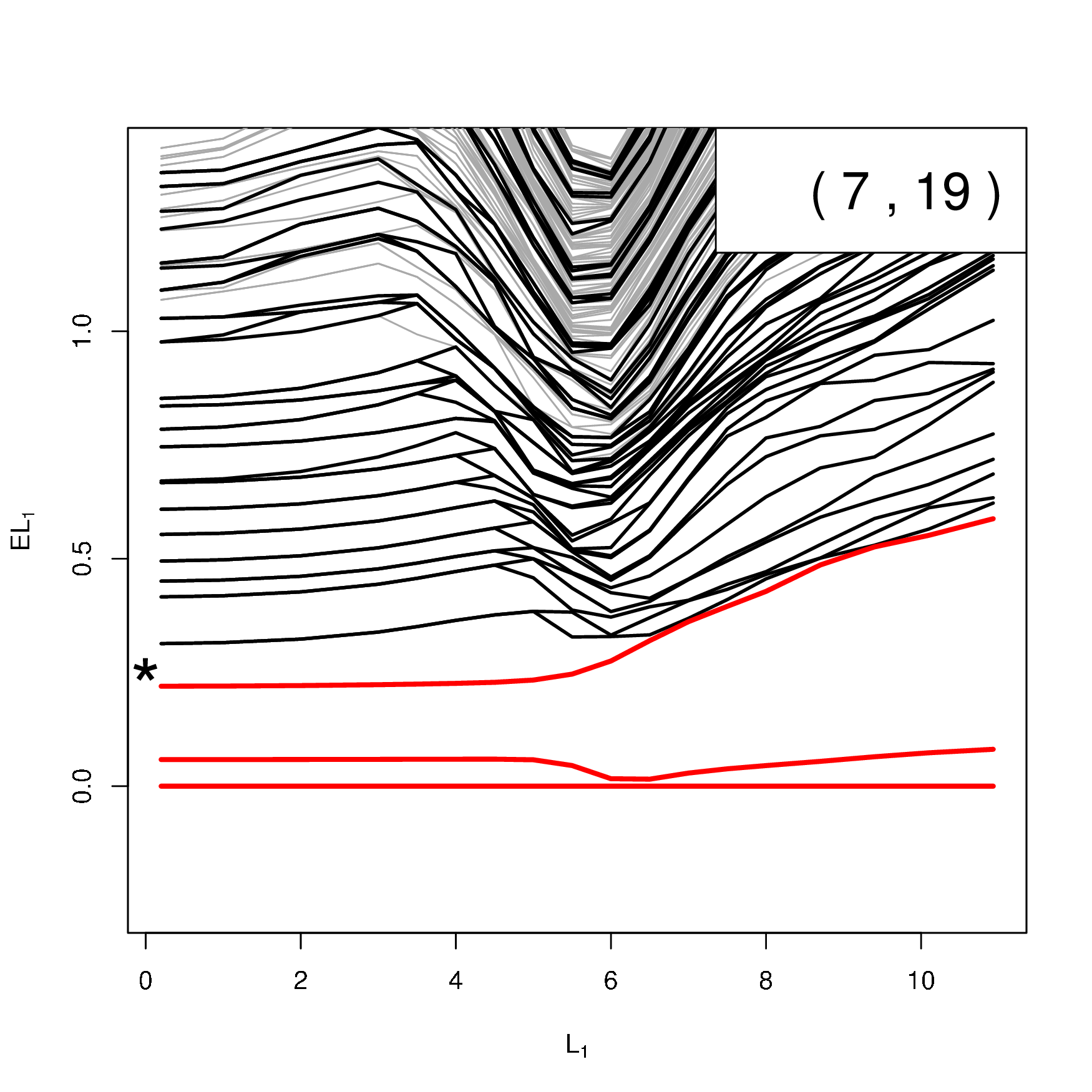}
\par\end{centering}

\caption{Energy spectra for one $(N,N_s)=(6,17)$ and two (7,19) particles at $\nu=1/3$ with five unit cells---these excitations, identified in the TT-limit, are shown as red lines. 
\label{fig:5_15_p_pp}}
\end{figure}

\begin{figure}
\begin{centering}
\includegraphics[width=0.90\columnwidth]{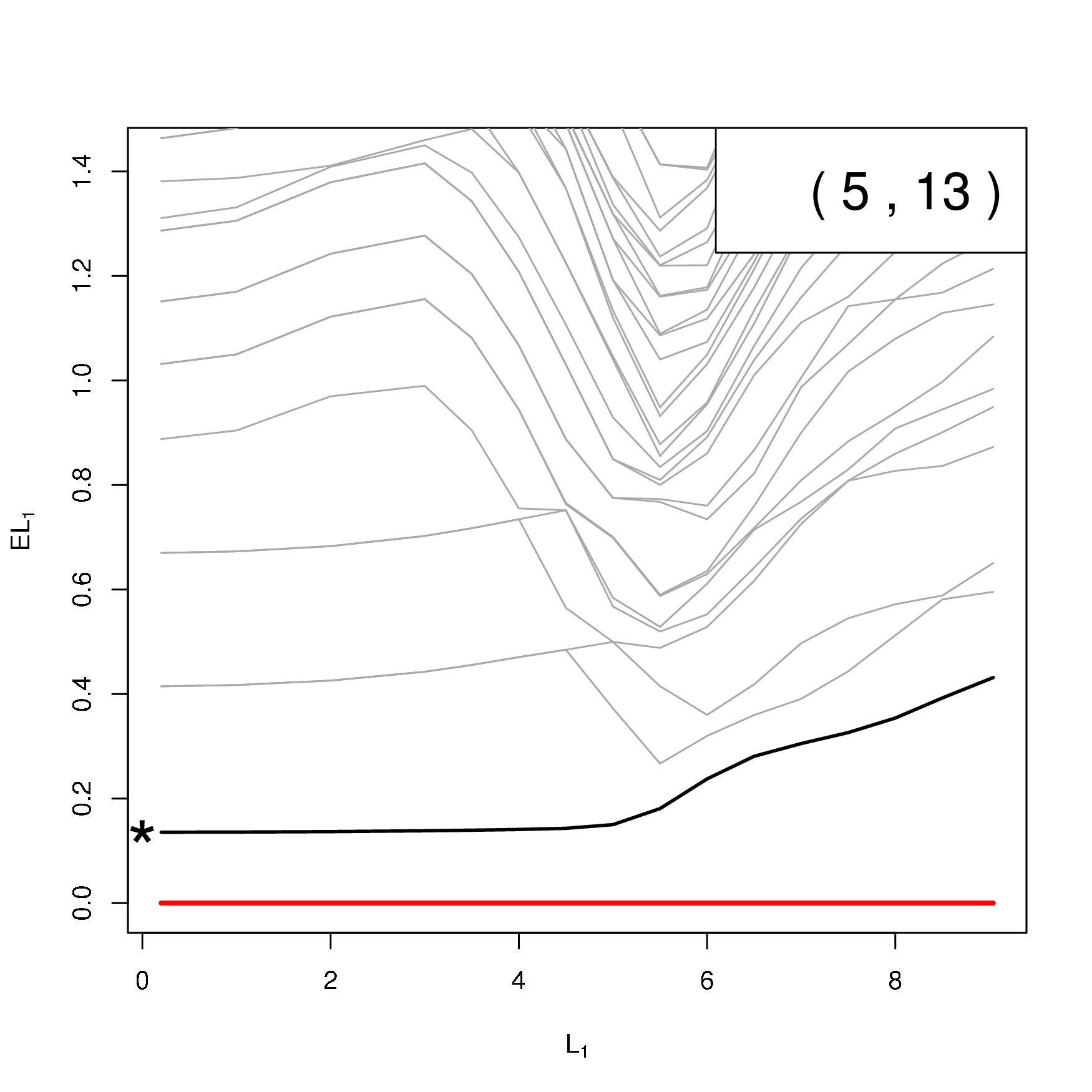}
\includegraphics[width=0.90\columnwidth]{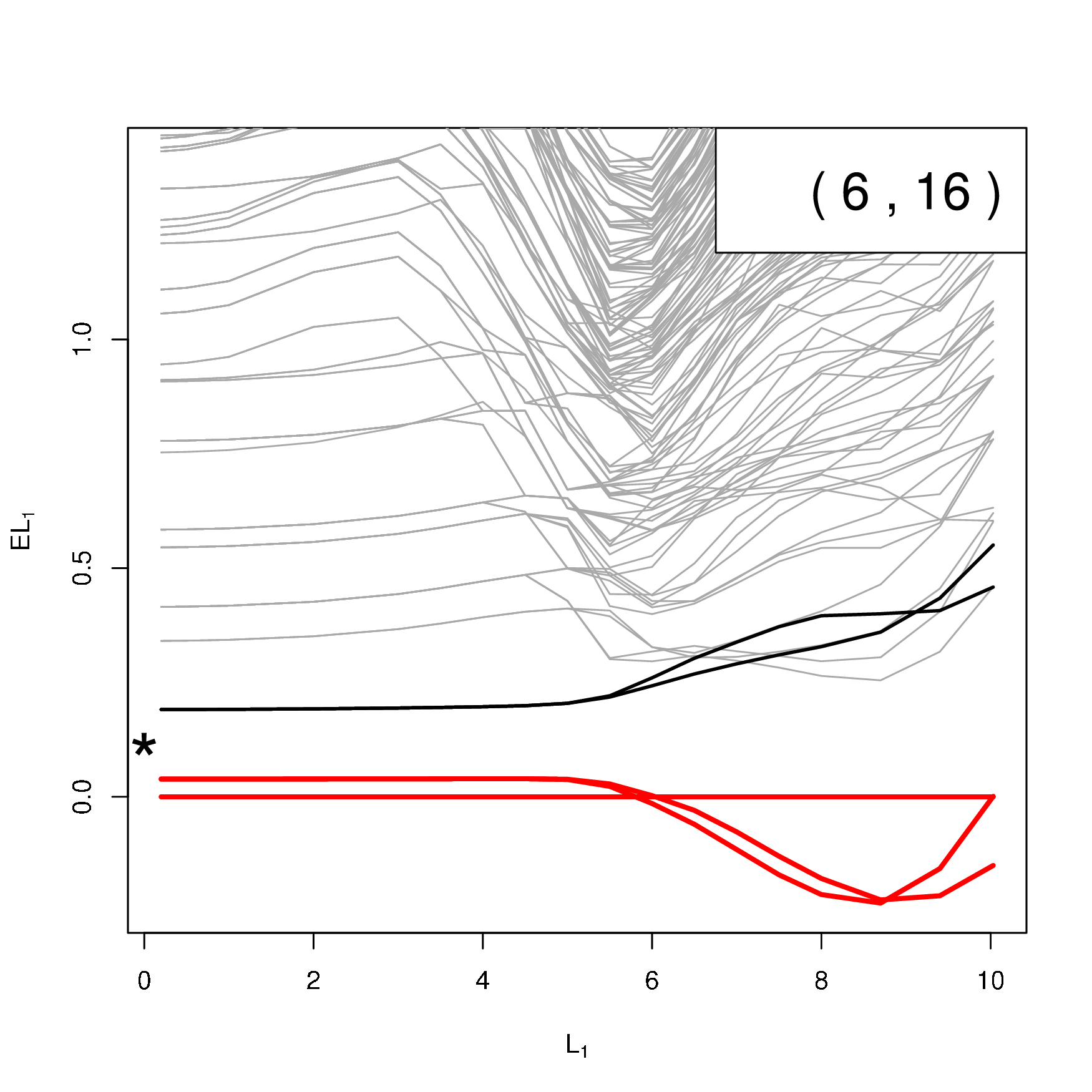}
\par\end{centering}

\caption{Energy spectra for one $(N,N_s)=(5,13)$ and two (6,16) holes at $\nu=2/5$ with two unit cells---these excitations, identified in the TT-limit, are shown as red lines. 
\label{fig:4_10_h_hh}}

\end{figure}

\begin{figure}
\begin{centering}
\includegraphics[width=0.90\columnwidth]{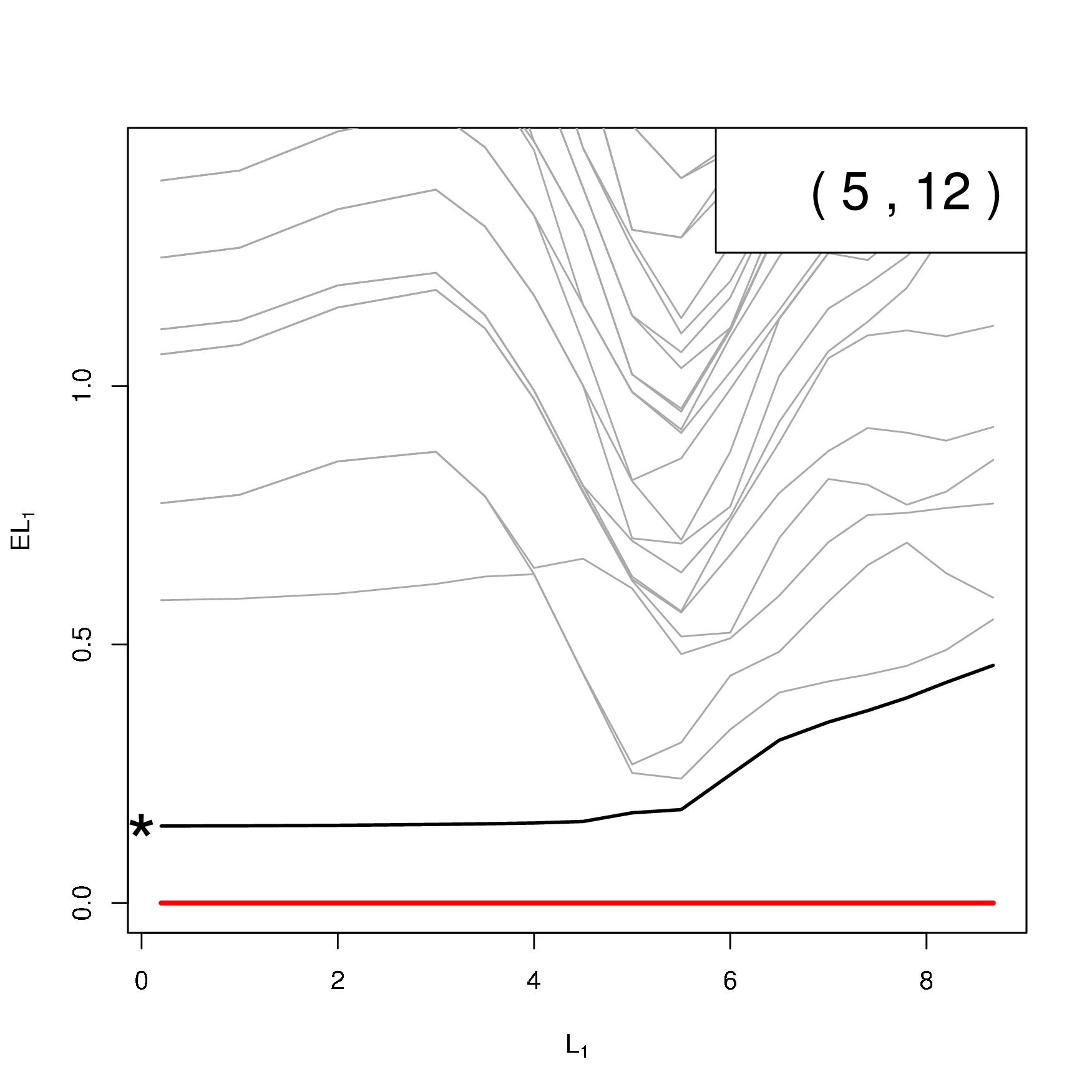}
\includegraphics[width=0.90\columnwidth]{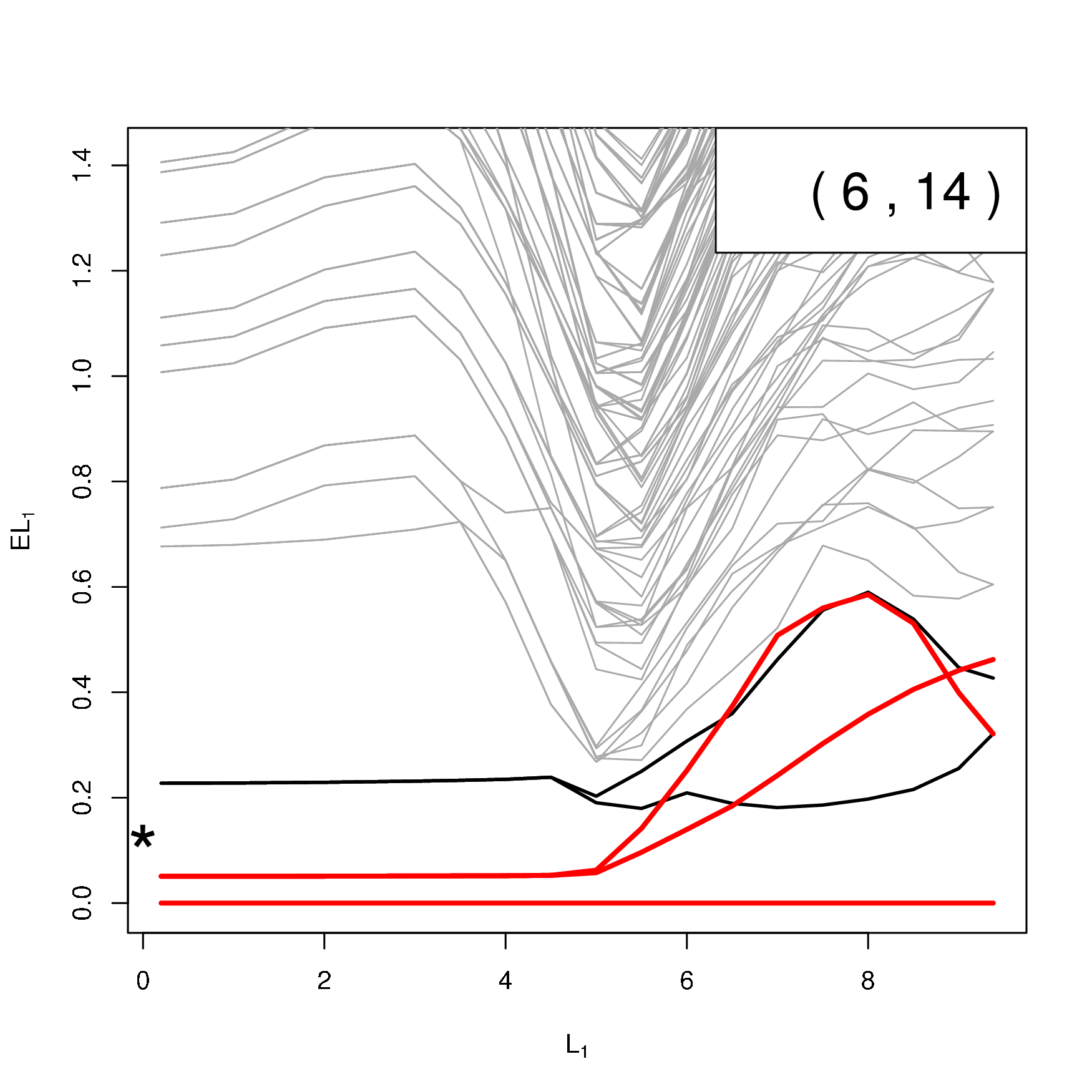}
\par\end{centering}

\caption{Energy spectra for one $(N,N_s)=(5,12)$ and two (6,14) particles at $\nu=2/5$ with two unit cells---these excitations, identified in the TT-limit, are shown as red lines.
\label{fig:4_10_p_pp}}

\end{figure}

Fig. \ref{fig:5_15_h_hh} shows the energy spectra for one and two holes in the 1/3 ground state, whereas Fig. \ref{fig:5_15_p_pp} shows the data for one and two particles.
The corresponding spectra for 2/5 are shown in Fig. \ref{fig:4_10_h_hh} and \ref{fig:4_10_p_pp}.
We see that the one hole, as well as the one particle, states are the lowest energy states for all $L_1$ at both 1/3 and 2/5.
The same holds true for the two hole states, however at 1/3 the gap at small $L_1$ is small.
Thus for all these cases the lowest energy states identified in the TT-limit remain the low-energy states for all $L_1$.
For the two-quasiparticle excitations the situation is different.
At 1/3 one of the low energy states identified in the TT-limit disappears to higher energies as $L_1$ increases.
Similarly, at 2/5 the low energy states also change.
We conclude that the few particle and hole excitations identified in the TT-limit show a large, but not complete, correspondence to the states at large $L_1$.
In the next Section we analyze these excitations, and the relation between the TT-limit and the bulk, in terms of exclusion statistics.

In all the figures something happens in the spectrum at $L_1 \approx 6$, this is where the hopping matrix elements $V_{km}, \, m\neq 0$ become important.
In Fig. \ref{fig:4_10_h_hh}, for $(N,N_s)=(6,16)$, there is a change of ground state.
Note that this describes the ground state at 3/8 as well as two holes at 2/5.
The change in ground state as $L_1$ increases is typical for even denominator filling factors \cite{bk2}.
The finite size dependence of the crossover caused by the onset of hopping, has been carefully studied at $\nu=1/2$ where it was found that a transition occurs at $L_1=5.3$ virtually independent of $N_s$ \cite{emil}.
This agrees with the very weak dependence of size we see here and we conclude that the crossover remains at finite $L_1$ as $N_s$ increases.

We have verified that the qualitative features obtained above are unchanged if the unscreened Coulomb potential is replaced by a screened one or by a short-range, pseudo-potential, interaction.

\section{Exclusion statistics}\label{exclusion}

The low energy quasiparticle states can be analyzed in terms of the exclusion statistics introduced by Haldane \cite{exclusion}.
The exclusion statistics parameters $g_{\alpha \beta}$ are defined as
\begin{eqnarray}\label{def}
\Delta d_\alpha = -\sum_\beta g_{\alpha \beta}\Delta N_\beta \ ,
\end{eqnarray}
where $\Delta d_\alpha $ is the change in the number of states, $d_\alpha$,
available to a particle of species $\alpha$ when the number, $N_\beta$,
of particles of species $\beta$ is changed by $\Delta N_\beta$ (keeping the size of the system unchanged).  
By exact diagonalization of small systems on the sphere and by using the composite fermion construction,
the following values have been established $g_{++}=g_{-+}=-g_{+-}=2-g_{--}=1+2n/(2np+1)$  for the Jain series $\nu=p/(2np+1)$\cite{canright,wu94,su,isakov}.
(Except that Ref. \onlinecite{wu94} claims that $g_{-+}=-g_{+-}=g_{--}$.)
(The diagonal parameters were determined  for any level of hierarchy states in Ref. \onlinecite{sitko03}.)
Since we identified in Sec. \ref{Low_section} the low energy states in the TT-limit as particle and hole excitations it is straightforward to determine the exclusion statistics in this limit for any $\nu=p/q$,
see \ref{app:exclusion}. 
This gives  $g_{\mf{pp}}=\frac{l_{\mf{p}}}{q}$,
$g_{\mf{hp}}=-\frac{l_{\mf h}}{q}$, $g_{\mf{h}\mf{h}}=\frac{l_{\mf{h}}}{q}$ and $g_{\mf{p}\mf{h}}=-\frac{l_{\mf{p}}}{q}$,
where the lengths $l_\pm$ of the quasiparticles are given by \pref{ln}.
This differs from the established values, in particular we note that the TT-limit gives $g_{--}+g_{++}=1$ rather than $g_{--}+g_{++}=2$, which, 
 at $\nu=1/q$, is obtained by assuming the quasiparticles are anyons and counting the number of anyon wave functions \cite{wu94,canright94,sitko03}.
At 1/3 the TT-limit gives $g_{--}=1/3$, $g_{++}=2/3$ as compared to $g_{--}=1/3$, $g_{++}=5/3$,
whereas at 2/5 the values are  $g_{--}=3/5$, $g_{++}=2/5$ and  $g_{--}=3/5$, $g_{++}=7/5$ respectively.
We see that the values for the quasiholes agree, 
whereas those for the quasielectrons are shifted by one. 
 
This is just a rephrasing in terms of exclusion statistics of the results we found in Sec. \ref{Numerics}. In the figures above, the statistics parameters in the TT-limit can be read off from the red states, 
which are the one and two particle/hole states in this limit, see \ref{app:exclusion}. The exclusion statistics is unchanged as $L_1$ increases if and only if the red states remain the low energy states. 
We see that this is the case for two holes at 1/3 and 2/5, {\it cf} Fig. \ref{fig:5_15_h_hh} , \ref{fig:4_10_h_hh}, whereas red states disappear from the low energy spectrum as $L_1$ increases for two particles, {\it cf} Fig. \ref{fig:5_15_p_pp}, \ref{fig:4_10_p_pp}.
We conclude that the exclusion statistics may change when $L_1$ increases from the TT-limit to the physical regime.
 
The non-diagonal exclusion parameters, $g_{-+},g_{+-}$, can be inferred from Fig. \ref{fig:4_10_ph_5_15_ph}, where the red lines give these parameters.
When $L_1$ increases these states mix with others, indicating that mutual statistics of particles and holes changes.
However,  there is no obvious separation of a low energy sector at $L_1=L_2$  and a value for the off-diagonal components cannot be extracted from these small systems.

\section{Discussion}\label{discussion}

Earlier studies have shown that the ground state in the Tao-Thouless limit, at any filling fraction $\nu=p/q\le 1$,  is adiabatically connected to an abelian quantum Hall state in the two-dimensional bulk system and the fractionally charged quasiparticles, $\mf p, \, \mf h$, are domain walls in the TT-limit.
In this article we have studied the full low-energy sector of states in the TT-limit.
Using the fact that the unit cell, $C$, that gives the ground state is $C=\mf h \mf p$ we found that, assuming a certain condition for the interaction ($V''_k \gg V''_{k+1}$), the low-energy sector consists of the reorderings of the $\mf p$ and $\mf h$ in the string of $C$'s that make up the ground state; these reorderings correspond to (multiple) particle-hole excitations.
From the low-energy sector we determined the exclusion statistics in the TT-limit.

\begin{figure}
\begin{centering}
\includegraphics[width=0.90\columnwidth]{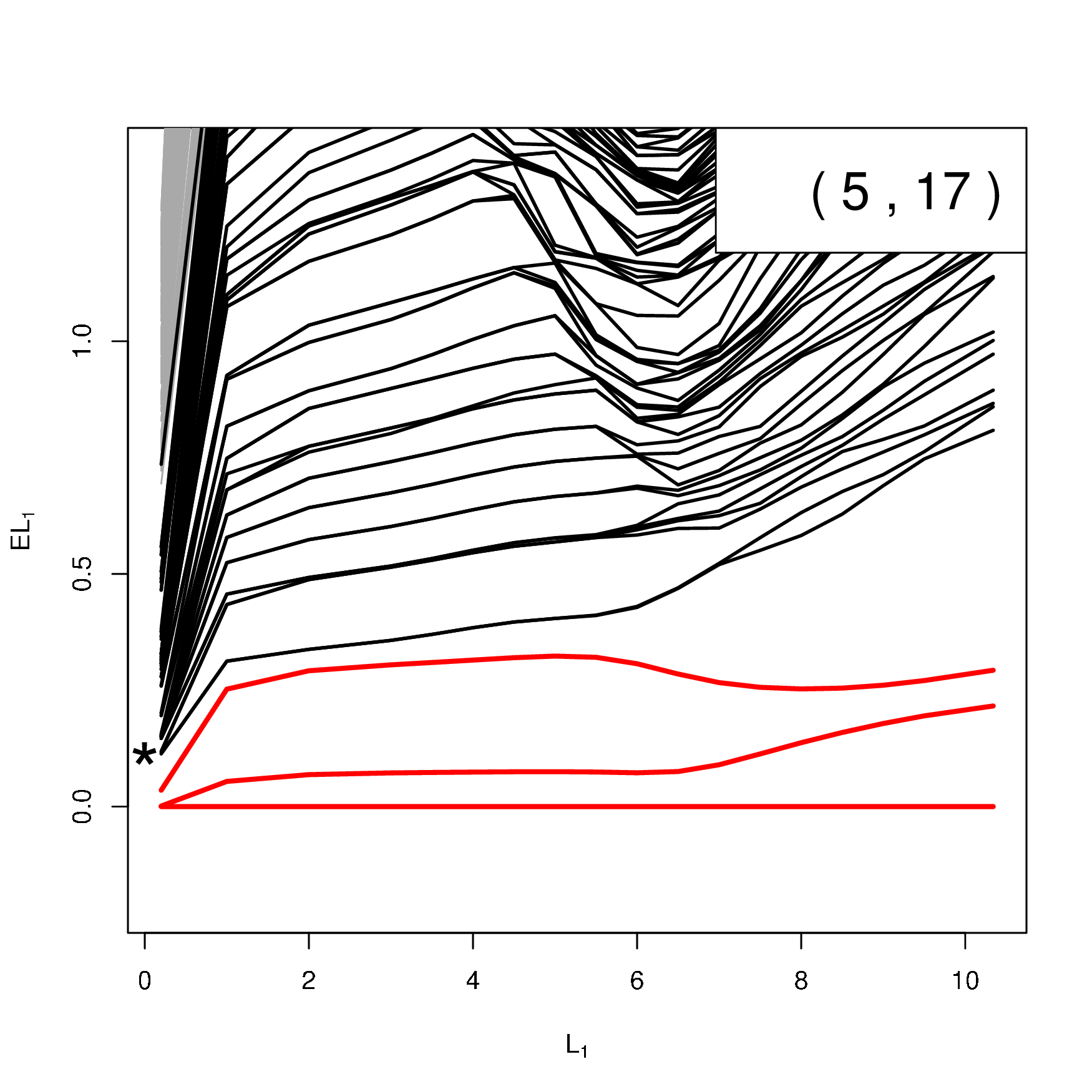}
\includegraphics[width=0.90\columnwidth]{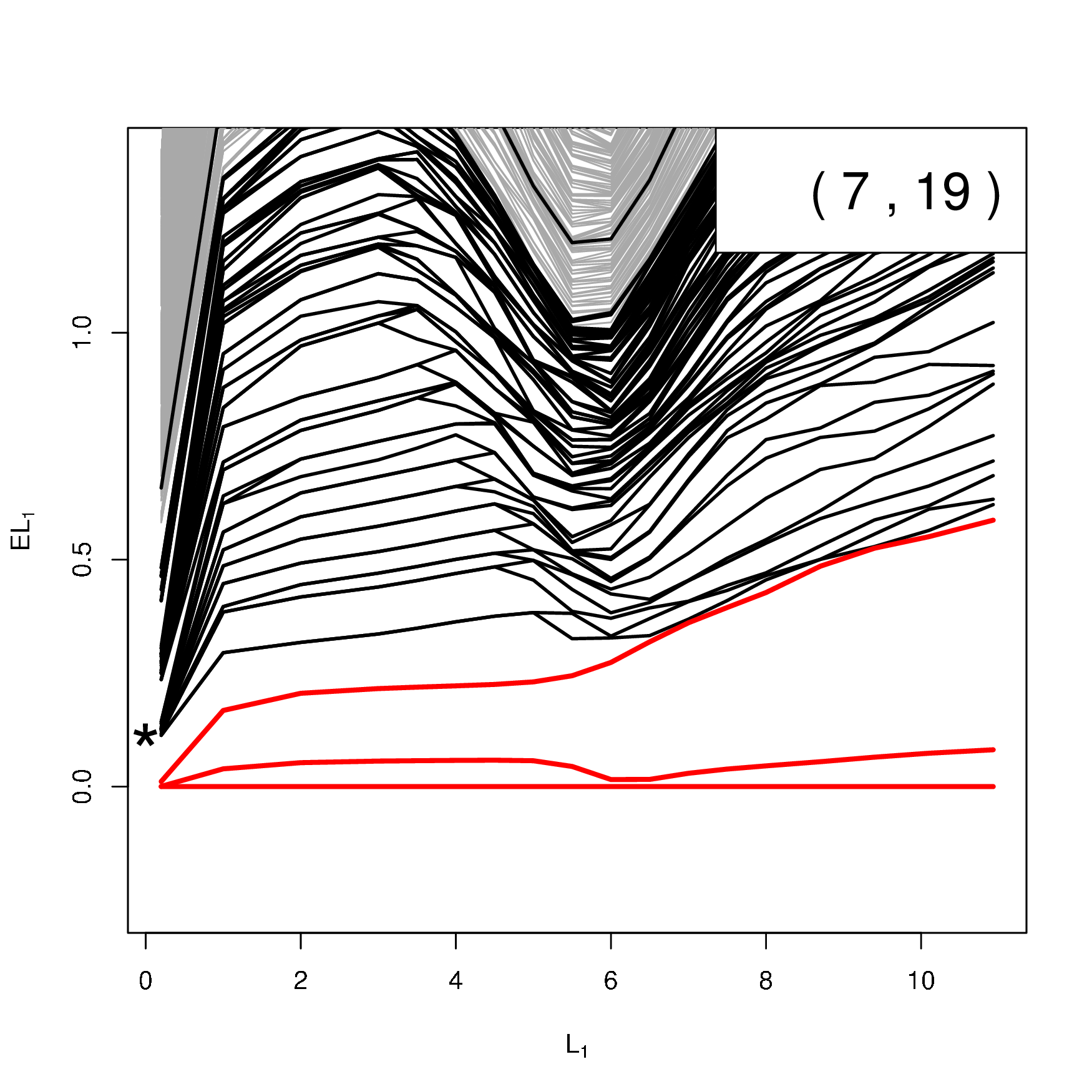}
\par\end{centering}

\caption{Energy spectra with screened Coulomb interaction for two $(N,N_s)=(5,17)$ quasiholes and two (7,19) quasiparticles at $\nu=1/3$. The low-energy states are clearly separated in the TT-limit which is not the case for the unscreened Coulomb interaction in Fig.  \ref{fig:5_15_h_hh} and \ref{fig:5_15_p_pp}. 
  \label{fig:5_17_yuk}}
\end{figure}

We studied numerically, at and around $\nu=1/3$ and 2/5, how the low-energy states develop as one moves away from the TT-limit.
We found that the low energy states in the physical regime always develop from low-energy TT-states.
This is a non-trivial result which supports the TT-approach.
However, there is not a perfect match between the lowest energy states in the TT-limit and those in the physical regime and this lack of matching is reflected in different exclusion statistics in the two regimes.
The states that give the TT-exclusion statistics at $\nu=1/3$ and 2/5 are the states marked in red in the figures above.
Consider for example two holes and two particles at $\nu=1/3$, Fig. \ref{fig:5_15_h_hh} and \ref{fig:5_15_p_pp}.
The two-hole states are low-energy states for all $L_1$, whereas one of the two-particle states disappears to higher energies as $L_1$ increases.
However, although the red states are the lowest energy states in the TT-limit, $L_1\rightarrow 0$, the gap to higher energy states is not big, and is certainly not diverging as $L_1\rightarrow 0$.
This is because the Coulomb interaction obeys only  $V_k> V_{k+1}$, with a small difference between successive terms.
From just considering the energies in the TT-limit, one may question if the red states are really the low-energy states.
If instead a screened Coulomb interaction is used, see Fig. \ref{fig:5_17_yuk}, then the red states are clearly separated as the low-energy states.
This indicates that the precise identification of the very lowest energy states in the TT-limit depends on the details of the interaction.
Note that there is no unique way of taking the limit of, say,  the Coulomb interaction as $L_1$ decreases; since this is an unphysical process one may allow the interaction to change.
The TT-ground state is insensitive to details of the interaction (as long as $V''_k>0$); it is not surprising that this is not the case for the gaps between the low-energy excitations.
It is conceivable that one could modify the interaction, for small $L_1$, so that the exclusion statistics in the TT-limit agrees with the one in the physical regime.
However, we find this unnatural since the counting of the low-energy states would then no longer be the counting of the domain walls in the TT-limit.
Thus we conclude that the exclusion statistics does change as $L_1$ increases.
Still, one should keep in mind that there is a large correspondence between the lowest energy states for large and small $L_1$; all low energy states in the physical regime emanate from low energy states in the TT-limit.
This supports the validity of the TT-limit for the abelian quantum Hall states.

\begin{acknowledgments}
We would like to thank Emil J Bergholtz and Thors Hans Hansson for discussions.
This work was supported by Swedish Science Research Council.
\end{acknowledgments}

\appendix

\section{Relaxation procedure}\label{app:ph-properties}

The relaxation scheme states that the positions of the electrons on the
sites are given by 
\begin{eqnarray}\label{relax}
x_{j}^{\left(\delta\right)}\equiv_{q}\left[\frac{j}{p}q+\delta\right] \ , 
\end{eqnarray}
where $j=1,\dots,p$, and $\left[\dots\right]$ is the rounding function
$\left[n+\epsilon\right]=n$ if $-\frac{1}{2}\leq\epsilon<\frac{1}{2}$.
We immediately see that 
\begin{eqnarray*}
x_{j}^{\left(\delta+q\right)}\equiv_{q}x_{j}^{\left(\delta\right)}+q\equiv_{q}x_{j}^{\left(\delta\right)} \\
x_{j+p}^{\left(\delta\right)}\equiv_{q}x_{j}^{\left(\delta\right)}+q\equiv_{q}x_{j}^{\left(\delta\right)}\ . 
\end{eqnarray*}
Thus we need only consider $0<j\leq p$ and $0\leq\delta<q$.\\

\textbf{Definition:} The ordered set $C_{\left(\delta\right)}=\left\{ x_{j}^{\left(\delta\right)},j\in1,\dots,p\right\} $
is the unit cell; the transposed set is defined as $C^T_{\left(\delta\right)}=\left\{ x_{p+1-j}^{\left(\delta\right)},j\in1,\dots,p\right\} $. 

The sets $C_{\left(\delta\right)}$, for different $\delta$, differ  by a translation, $x_j^{(\delta)}\rightarrow x_j^{(\delta)} + n$, at most.
Choosing different $\delta$, exactly $q$ different unit cells are obtained; these are related by rigid translations and correspond to the $q$-fold degeneracy on the torus.
The integers  $x_{j}^{\left(\delta\right)}$ change in discrete steps when $\delta$ changes. 
Specifically, the change to the  set $C_{\left(\delta\right)}$ under $\delta \rightarrow \delta + \frac{1}{p}$ is that one $x_j$ changes by one.
This hopping of one specific electron one site is thus equivalent to a translation of the unit cell.

Transposition of a set, $C^T_{\left(\delta\right)} $, is equivalent to a translation.
One way to see this is to consider the transposition of $C_R=C_{\frac 1 2 -\frac 1 {2p}}$, see below. For this set, transposition is equivalent to hopping a specific electron which is equivalent to a translation.
Noticing that any transposition can be divided into a translation to $C_R$, a transposition and a translation back again leads to the conclusion that all transpositions are translations.

Since all sets $C_{\left(\delta\right)}$ are related by rigid translations, it follows that when $\delta$ changes by $\frac{a}{p}$ there exists a translation
$l_{a}$ and a corresponding relabeling $n_{a}$ that has the
same effect: 
\begin{eqnarray}
\forall a\in\mathbb{Z},\;\exists l_{a},n_{a}\in\mathbb{Z}:x_{j}^{\left(\delta+\frac{a}{p}\right)}=x_{j+n_a}^{\left(\delta-l_a\right)} \ ,
\end{eqnarray}
or, equivalently,
\begin{eqnarray}\label{nleqn}
\left[\frac{j}{p}q+\delta+\frac{a}{p}\right]=\left[\frac{j+n_a}{p}q+\delta-l_a\right] \ . 
\end{eqnarray}
This relation is trivially true when
\begin{equation}
a=n_aq-l_ap \ . \label{eq:main-equation}
\end{equation}

We see that if $l_a$ and $n_a$ satisfy \eqref{eq:main-equation} then so do $l_a+q$ and $n_a+p$.
This gives the general solution to \eqref{nleqn}.
Since all the variables in \eqref{eq:main-equation} are integers the equation is a modulo equation of type 
\begin{eqnarray*}
l_ap & \equiv_q & -a\\
n_aq & \equiv_p & a \ . 
\end{eqnarray*} 
We solve these equations explicitly using Euler's theorem:
If $\gcd\left(a,p\right)=1$, then $1\equiv_p a^{\varphi\left(p\right)}$, 
where $\varphi\left(p\right)$ is the number of integers smaller than or equal to $p$ that are relatively prime to $p$.
This gives
\begin{eqnarray}
  l_a & \equiv_{q} & -ap^{\varphi\left(q\right)-1}\label{eq:l_a-explicit}\\
  n_a & \equiv_{p} & aq^{\varphi\left(p\right)-1}\label{eq:n_a-explicit} \ .
\end{eqnarray}
From \eqref{eq:l_a-explicit} and \eqref{eq:n_a-explicit} it is trivial to see that $l_a+l_{-a}\equiv_{q}q$ and $n_{a}+n_{-a}\equiv_{p}p$.

Thus moving one particular electron one step ($a=\pm1$) corresponds to translating the unit cell $l_{\pm 1}$ steps.
This defines a partition of the unit cell where one part consists of $l_1$ sites with $n_1$ electrons and the other part consists of  $l_{-1}$ sites with $n_{-1}$ electrons, $l_1+l_{-1}=p, \, n_1+n_{-1}=q$.
These parts have charges $n_{\pm 1} -l_{\pm 1} p/q=\pm 1/q$ relative to the ground state.
We denote the partitions with charges $+1/q$ and $-1/q$ by $\mf p$ and $\mf h$ respectively.
These are the charges of the elementary quasiparticles and since the parts obey the relaxation condition \eqref{relax} it follows that they are the quasiparticles.
Moving the electron one step interchanges $\mf p$ and $\mf h$ in the unit cell.

We verify that the partition corresponding to the quasiparticle obeys the relaxation condition by checking that $x_j^{(\delta)}$ given by \eqref{relax} for $j=1,\dots n_a$ are unchanged if $p,q$ are replaced by $n_a, l_a$.
Consider, to be definite, the relaxation procedure $x_j=\left[j \frac{q}{p}\right]_\mathrm{f}$, where $\left[\dots\right]_{\mathrm f}$ is rounding downwards ($\left[x\right]_{\mathrm f}=\left[x-\frac{1}{2}\right]$) and $j=0,\dots,p-1$.
This corresponds to the $C_L$ representation introduced below.
The partitions are then given by $x^{(a)}_j=\left[j \frac{l_a}{n_a}\right]_\mathrm{f}$, where $j=0,\dots,n_a-1$.
If $x_j=x_j^{(a)}$ for all $j=0,\dots,n_a-1$ then the relaxation procedure works for the partitions as well.
We have checked this condition for all $p$, $q$ with $1=\gcd\left(p,q\right)$ and $0<p<q<100$ but lack a general proof.

\begin{figure}
\begin{centering}
\includegraphics[width=1\columnwidth]{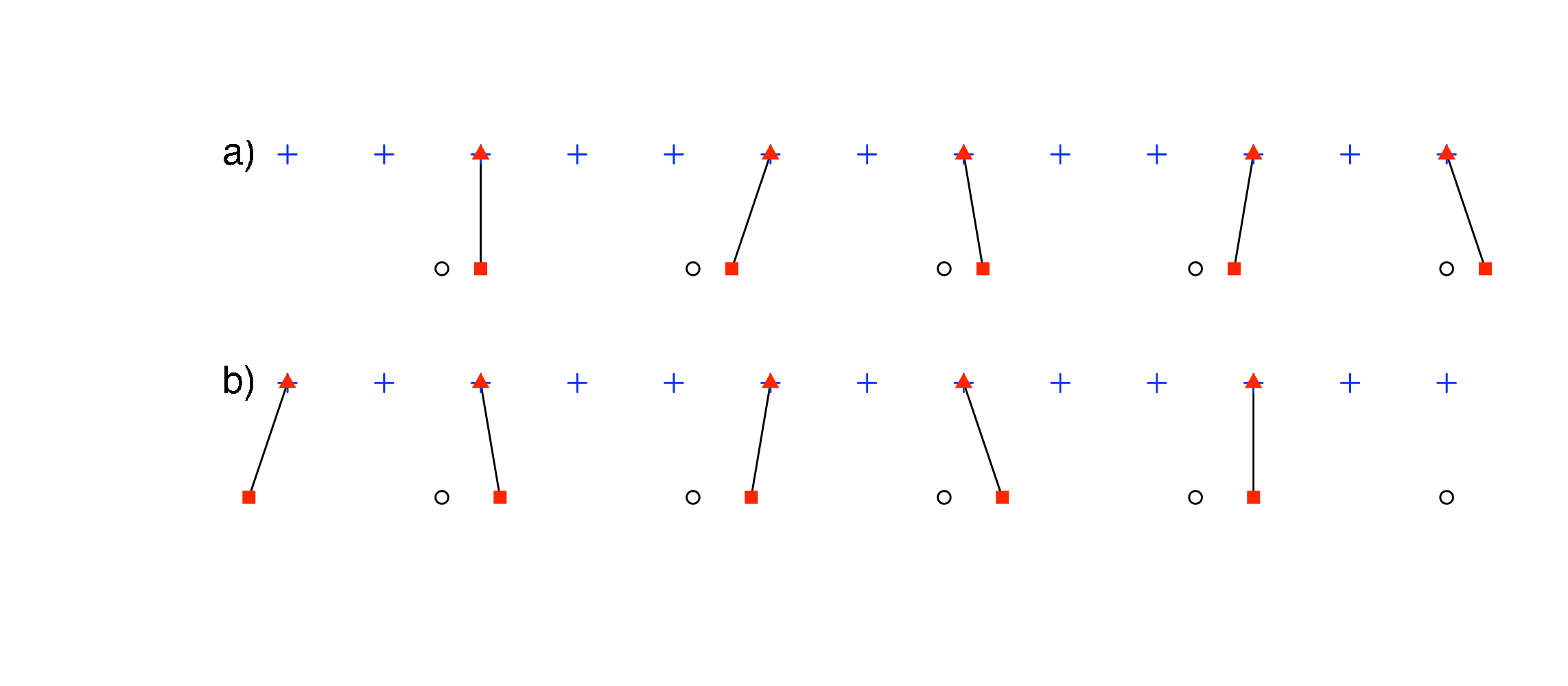}
\par\end{centering}
\caption{Relaxation scheme for (a) $C_R$,  $\delta=\frac{p-1}{2p}=\frac{2}{5}$, and (b) $C_L$, $\delta=\frac{p+1-2q}{2p}=-2$  at $\nu=\frac{5}{13}$.
Blue $+$ are positions of the sites. Black $\circ$ are unrelaxed electrons at $\delta=0$.
Red $\blacktriangle$ and $\blacksquare$ are relaxed and unrelaxed electrons for $C_R$ and $C_L$ respectively.
\label{fig:relax}}
\end{figure}

We now consider $\nu=5/13$ as an explicit example.
This illustrates the results in the Lemmas below, which follow by simple generalizations of the example. 
Fig. \ref{fig:relax} shows two constructions of unit cells by relaxation for different $\delta$.
In ($a$), $\delta= (1-1/5)/2=2/5$ and the unit cell is $C_R=0010010100101$; 
whereas in ($b$), $\delta= (1+1/5)/2=3/5$ and $C_L=1010010100100$.
($\delta$ is defined modulo $p/q$ because of the periodic boundary conditions.)
$C_R$  ($C_L$) has the property that  the electrons are as far to the right (left) as possible. 
The particle and hole at 5/13 are, as shown above, $\mf{p}=00101$ and $\mf{h}=00100101$ respectively (in the $R$-representation).
We see that $C_R$ ($C_L$) starts with a 0 (1) and ends with a 1 (0) while the interior is symmetric and the same for $C_R$ and $C_L$: $C_R=0\tilde C 1, \ C_L=1\tilde C 0$.
Consequently, $C_R$ and $C_L$ are mirror images of each other $C_L=C_R^T$.

\textbf{Definition:}
For general $\nu=p/q$, we define the left and right unit cells as $C_{L}=C_{\left(\frac{1}{2}-\frac{1}{2p}\right)}$ and $C_{R}=C_{\left(\frac{1}{2}+\frac{1}{2p}\right)}=C_{\left(\frac{1}{2}+\frac{1}{2p}-\frac{q}{p}\right)}$ respectively.

Alternatively, these cells can be defined as
\begin{eqnarray*}
  C_{L} & = & \left\{ C_{\left(\delta\right)}:\:\inf_{\delta}\sum_{j}x_{j}^{\left(\delta\right)}\right\} \\
  C_{R} & = & \left\{ C_{\left(\delta\right)}:\:\sup_{\delta}\sum_{j}x_{j}^{\left(\delta\right)}\right\} \ ,
\end{eqnarray*}
which makes explicit that the cell is translated so that  the electrons are as far to the left (right) as possible.

From the definition the following results, which hold for general $\nu=p/q$, follow: 
\begin{enumerate}
\item \label{mirror} $C^T_R=C_L$.
\item \label{hopping}$C_{L}=C_{R+\frac{1}{p}}$, {\it ie} hopping one electron (the last)
to the right in $C_{R}$ gives $C_{L}$.
\item \label{third}$C_{L}=1\tilde{C}0$ and  $C_{R}=0\tilde{C}1$.
\item \label{symmetry}$\tilde C^T=\tilde C$.
\end{enumerate}
Here, \pref{mirror} and \pref{hopping} follow directly from the definition; \pref{hopping} implies \pref{third} because the electron that hops when $C_R\rightarrow C_L$ is the electron at the edge;
\pref{third} then implies \pref{symmetry} because of \pref{mirror}.

\textbf{Definition:}
Let $\mathbb{U}=\left\{C_{(\delta)}\right\}$ be the set of all unit cells (including all translations for all $\nu=p/q$) and let $\mathbb{U}_R,\mathbb{U}_L\subset\mathbb{U}$,
where $\mathbb{U}_R\cap\mathbb{U}_L=\emptyset$,
be the subsets of left and right unit cells, $\mathbb{U}_R=\left\{C_R\right\}$,
$\mathbb{U}_L=\left\{C_L\right\}$.

\textbf{Lemma:}
$\forall C\in\mathbb{U}\setminus\mathbb{U}_L\:\exists \mf{p},\mf{h}\in\mathbb{U}$ such that
$C=\mf{h}\mf{p}$ and $\forall C\in\mathbb{U}\setminus\mathbb{U}_R\:\exists \mf{p},\mf{h}\in\mathbb{U}$ 
such that $C=\mf{p}\mf{h}$ where $\mf p$ is the particle and $\mf h$ is the hole.
When $C\in\mathbb{U}_R$ then $\mf{p},\mf{h}\in\mathbb{U}_R$ and when $C\in\mathbb{U}_L$ then $\mf{p},\mf{h}\in\mathbb{U}_L$.

\emph{Proof:} 
We know that any $\nu=p/q$ has a particle-hole decomposition according to \pref{eq:main-equation}.
We see that we can rewrite \pref{eq:main-equation} as 
\begin{eqnarray*}
\frac{p}{q} 
= \frac{1}{q}\left( \frac{n_a}{l_a}\left(q-a\right) + \frac{n_a-1}{l_a}a\right) \\
= \frac{1}{q}\left( \frac{n_a}{l_a}\left(q+a\right) - \frac{n_a+1}{l_a}a\right) \ .
\end{eqnarray*}

The interpretation of the above equations is as follows. If one considers the $q$ unit cells of $\nu$
 and takes the mean density of the $l_a$ leftmost (or rightmost) sites one gets the mean density of $\nu=p/q$, see the example in Table \ref{tab:Cells-2_5}.
Denote this segment of $C$ by $L$ and its complement by $R$ so that $C=LR$.
We see that in $q-\left|a\right|$ cases the density of $L$ is $\frac{n_a}{l_a}$, whereas in the $\left|a\right|$ remaining $L$ the density is $\frac{n_a-sign(a)}{l_a}$.
For $a=1$ ($a=-1$) $L$ has the correct size and density to be the quasiparticle $\mf{p}$ (quasihole $\mf{h}$) discussed earlier.
Because of \pref{eq:main-equation}, $R$ has to be the quasihole $\mf{h}$ (quasiparticle $\mf{p}$).
The unit cell where $L\neq\mf{p}$ ($L\neq\mf{h}$) has to be $C_R$ ($C_L$) because in that cell the electrons are as far to the right (left) as possible. 

We can, because of the definition of $\mathbb{U}_R$ ($\mathbb{U}_L$), conclude that by necessity $\mf{h}_R,\mf{p}_R\in\mathbb{U}_R$ ($\mf{h}_L,\mf{p}_L\in\mathbb{U}_L$).
\begin{flushright}$\square$\par\end{flushright}

\begin{table}
  \begin{tabular}{|c|c|c|}\hline 
    $C$ & $a=-1$  & $a=1$
  \tabularnewline\hline\hline
    $C_R$ & $001\;01$ & $00\;101$
    \tabularnewline\hline 
    & $010\;01$ & $01\;001$
    \tabularnewline\hline 
    & $010\;10$ & $01\;010$
    \tabularnewline\hline 
    & $100\;10$ & $10\;010$
    \tabularnewline\hline 
    $C_L$ & $101\;00$ & $10\;100$
    \tabularnewline\hline 
  \end{tabular}
  \caption{The five representations of the unit cell at $\nu=2/5$. Notice how hopping one electron one site causes the $\mf p$ and $\mf h$ parts to trade places.\label{tab:Cells-2_5}}
\end{table}

Thus, as Table \ref{tab:Cells-2_5} gives an example of for $\nu=\frac{2}{5}$, we can decompose the unit cells  into the corresponding quasiparticles and quasiholes: $C_{L}=\mf{p}_{L}\mf{h}_{L}$ and $C_{R}=\mf{h}_{R}\mf p_R$. 
(The order of $\mf{p}$ and $\mf{h}$ is given by the fact that  $\mf{p}$ has a higher density of electrons than  $\mf{h}$, {\it ie}  a larger filling fraction, and that 
$C_L$ has the electrons as far to the left as possible.)  $\mf{p}$ and  $\mf{h}$ are themselves unit cells defined by the relaxation procedure at appropriate filling factors,
consequently, 
\begin{enumerate}\setcounter{enumi}{4}
\item $\mf{p}_{L}^T=\mf p_R$, $\mf{h}_{L}^T=\mf{h}_{R}$\label{enu:transposition}
\end{enumerate}
(alternatively, this follows from $C^T_L=C_R$). One readily finds that
\begin{enumerate}\setcounter{enumi}{5}
\item $\mf{p}_{L}=1\tilde{\mf{p}}_{L}$, $\mf{h}_{L}=\tilde{\mf{h}}_{L}0$;\\
  $\mf p_R=\tilde{\mf{p}}_{R}1$, $\mf{h}_{R}=0\tilde{\mf{h}}_{R}$;\\
  $\tilde{\mf{p}}_{L}^T=\tilde{\mf p}_R$, $\tilde{\mf{h}}_{L}^T=\tilde{\mf{h}}_{R}$ \label{enu:def-tilde}.
\item $\tilde{\mf{p}}_{L}\tilde{\mf{h}}_{L}=\tilde{\mf{h}}_{R}\tilde{\mf{p}}_{R}=\tilde{C}$. \label{enu:mirrors}
\end{enumerate}

\textbf{Lemma:}
If $C_{R}=\mf{h}_{R}\mf p_R\in\mathbb{U}_R$ then $\mf{h}_{R}^n\mf p_R\in\mathbb{U}_R$, where $\mf{h}_{R}^{n-1}\mf p_R\in\mathbb{U}_R$  is the quasiparticle
(and $\mf{h}_{R}$ is the quasihole), and $\mf{h}_{R}\mf p_R^m\in\mathbb{U}_R$, where $\mf{h}_{R}\mf p_R^{m-1}\in\mathbb{U}_R$ is the quasihole  (and $\mf p_R$ is the quasiparticle), for $n,m\in\mathbb{Z}$.

\emph{Proof:}
We prove that $\mf h_R\mf p_R^2\in\mathbb U_R$ with $\mf h_R\mf p_R$ as hole and $\mf p_R$ as particle. The general result then follows by induction.
Rewriting \eqref{eq:main-equation} for $a=1\sim\mf p$ as 
\begin{eqnarray*}
1  & = & n_{\mf p} q - l_{\mf p} p \\
& = & n_{\mf p} (q+l_{\mf p}) -  l_{\mf p} (p+n_{\mf p}) \ ,
\end{eqnarray*}
we see that if $l_{\mf p}$ and $n_{\mf p}$ solve \eqref{eq:main-equation} with $\nu=p/q$ they also solve it for
$\nu=\frac{p+n_{\mf p}}{q+l_{\mf p}}$.
This means that if one has one unit cell $C$ and adds an extra particle section then the quasiparticle of that new unit cell will be unchanged.
By induction this construction can be extended to any $\mf h_R\mf p_R^n$.
The proof for $\mf h_R^m\mf p_R$ is completely analogous.
\begin{flushright}$\square$\par\end{flushright}

\textbf{Lemma:}
If $C_{R}=\mf{h}_{R}\mf p_R\in\mathbb{U}_R$, then 
\begin{eqnarray}
C_R & = & 
\begin{cases}
\mf{h}_R^{n+1}\mf{p}_R^\prime &
\mbox{if }l\left(\mf{h}_R\right)<l\left(\mf{p}_R\right)\\
\mf{h}_R^\prime\mf{p}_R^{m+1} & 
\mbox{if }l\left(\mf{h}_R\right)>l\left(\mf{p}_R\right) \ ,
\end{cases}\label{eq:repetetive_terms}
\end{eqnarray}
where $l\left(\mf{h}_R\right)\geq l\left(\mf{p}_R^\prime\right)$ or $l\left(\mf{h}_R^\prime\right)\leq l\left(\mf{p}_R\right)$ respectively.
Equality holds when $\mf{p}_R^\prime=1$ or $\mf{h}_R^\prime=0$. 
Here $\mf{p}_R^\prime$  is the quasiparticle in the unit cell $\mf{h}_R\mf{p}_R^\prime$ (and $\mf{h}_R^\prime$ is the quasihole in $\mf{h}_R^\prime\mf{p}_R$).

\emph{Proof:}
We prove that $\mf h_R\mf p_R=\mf h^\prime_R\mf p_R\mf p_R$ if $l\left(\mf{h}_R\right) > l\left(\mf{p}_R\right)$.
Rewriting \eqref{eq:main-equation} for $a=1\sim\mf p$ as 
\begin{eqnarray*}
1  & = & n_{\mf p} q - l_{\mf p} p \\
& = & n_{\mf p} (q-l_{\mf p}) -  l_{\mf p} (p-n_{\mf p}) \ ,
\end{eqnarray*}
we see that if $l_{\mf p}$ and $n_{\mf p}$ solve \eqref{eq:main-equation} with $\nu=p/q$ they also solve it for $\nu=\frac{p-n_{\mf p}}{q-l_{\mf p}}$.
In analogy with the preceding lemma the particle of a unit cell $C$ remains unchanged even if we remove a particle partition.
Recursive usage of this argument leads to \eqref{eq:repetetive_terms}.
The case of $l\left(\mf p_R\right)=l\left(\mf h_R\right)$ only occurs when $\nu=1/2$ which is trivial since $\mf p=1$, $\mf h=0$ and $n=m=0$.
\begin{flushright}$\square$\par\end{flushright}

This lemma also tells us how to decompose a state obtained by the relaxation procedure \eqref{relax}.
Since $\mf{h}_{R}$ and $\mf p_R$ are unit cells themselves the construction holds recursively.
The lemma suggests the following indexation that treats particles and holes on equal footing:
\begin{eqnarray}
 \mf{p}_{kR} & = & \mf{h}_{k+\frac{1}{2}R}^{n_k}\mf{p}_{k+1R}\label{eq:p_R_decompose} \\
 \mf{h}_{rR}& = &  \mf{h}_{r+1R}\mf{p}_{r+\frac{1}{2}R}^{m_r}\label{eq:h_R_decompose}  \ .
\end{eqnarray}
(Transposing gives the corresponding relations for the $L$-representation.)
These (half) integer indices are only a convenient way of numbering the different particles and holes and carry no deeper meaning.
We define the starting point for the indices by writing the particle-hole decomposition of the chosen, arbitrary, unit cell $C_R$ as 
\begin{eqnarray}
C_{R} & = & \begin{cases}
\mf{h}_{\frac{1}{2}R}\mf{p}_{0R} & \mbox{if }l\left(\mf{h}_{R}\right)\leq l\left(\mf p_R\right)\\
\mf{h}_{0R}\mf{p}_{\frac{1}{2}R} & \mbox{if }l\left(\mf{h}_{R}\right)\geq l\left(\mf p_R\right)\ . \end{cases}\label{eq:indexed_terms} \
\end{eqnarray}
Note that the particle and its corresponding hole have different indices.  Using this indexation we can now rewrite  \eqref{eq:repetetive_terms} as
\begin{eqnarray}
C_{R} & = & \begin{cases}
\mf{h}_{\frac{1}{2}R}^{n+1}\mf{p}_{1R} & \mbox{if }l\left(\mf{h}_{R}\right)\leq l\left(\mf p_R\right)\\
\mf{h}_{1R}\mf{p}_{\frac{1}{2}R}^{m+1} & \mbox{if }l\left(\mf{h}_{R}\right)\geq l\left(\mf p_R\right) \ . \end{cases}\label{eq:re_indexed_terms}
\end{eqnarray}

As an example of  our notation we consider the decomposition of $\nu=5/13$, where $C_R=\mf{h}\mf{p}$ with $\mf{h}=00100101$ and $\mf{p}=00101$, see Table \ref{tab:Hirearchy-5_13}. Since the 
hole is longer than the particle we have  $C_R=\mf{h}\mf{p}=\mf{h}_{0R}\mf{p}_{\frac{1}{2}R}$.
 $\mf{h}_{0R}$ can then be decomposed as $\mf{h}_{0R}=\mf{h}_{1R}\mf{p}_{\frac{1}{2}R}$ and, continuing the process,
$\mf{p}_{\frac{1}{2}R}=\mf{h}_{1R}\mf{p}_{\frac{3}{2}R}$,
$\mf{h}_{1R}=\mf{h}_{2R}\mf{p}_{\frac{3}{2}R}$, $\mf{p}_{\frac{3}{2}R}=\mf{h}_{2R}\mf{p}_{\frac{5}{2}R}$. Here $\mf{h}_{1R}=001$, $\mf{p}_{\frac{3}{2}R}=01$, $\mf{h}_{2R}=0$, $\mf{p}_{\frac{5}{2}R}=1$. 
Note how holes and particles are alternating at every step and that it is the one with the smallest index,
which is the longest one, that is decomposed.
In the example at hand $m_k=n_r=1$, {\it cf} Eq'n (\ref{eq:p_R_decompose}, \ref{eq:h_R_decompose}), this need of course not be the case. 

\begin{table}
  \begin{tabular}{|c|c|c|}\hline 
    $01$ representation & $\mf{ph}$ reprepresentation & level
    \tabularnewline\hline\hline
    $\left(00100101\right)\left(00101\right)$ 
    & $\mf{h}_{0R}\mf{p}_{\frac{1}{2}R}$
    & $\frac{5}{13}$
    \tabularnewline\hline 
    $\left(001\right)\left(00101\right)^2$ 
    & $\mf{h}_{1R}\mf{p}_{\frac{1}{2}R}^2$
    & $\frac{3}{8}$
    \tabularnewline\hline 
    $\left(001\right)\left(\left(001\right)\left(01\right)\right)^2$ 
    & $\mf{h}_{1R}\left(\mf{h}_{1R}\mf{p}_{\frac{3}{2}R}\right)^2$
    & $\frac{2}{5}$
    \tabularnewline\hline 
    $\left(0\right)\left(01\right)\left(\left(0\right)\left(01\right)^2\right)^2$
    & $\mf{h}_{2R}\mf{p}_{\frac{3}{2}R}\left(\mf{h}_{2R}\mf{p}_{\frac{3}{2}R}^2\right)^2$
    & $\frac{1}{3}$
    \tabularnewline\hline 
    $\left(0\right)^2\left(1\right)\left(\left(0\right)\left(\left(0\right)\left(1\right)\right)^2\right)^2$
    & $\mf{h}_{2R}^2\mf{p}_{\frac{5}{2}R}\left(\mf{h}_{2R}\left(\mf{h}_{2R}
    \mf{p}_{\frac{3}{2}R}\right)^2\right)^2$
    & $\frac{1}{2}$
    \tabularnewline\hline 
  \end{tabular}
  \caption{Hierarchy levels for $\frac{5}{13}$ with $C_{R}=0010010100101$. At each step the particles and holes correspond to a distinct value of $\nu$. Notice how for each step $\mf p$ and $\mf h$ are alternatingly expanded. \label{tab:Hirearchy-5_13}}
\end{table}

Property \ref{enu:transposition} above has as a consequence that all of the expansions (\ref{eq:repetetive_terms} - \ref{eq:re_indexed_terms}) have transposed variants.
The same holds true for the lemmas in the following section.

\section{Low energy states}\label{app:ph-proofs}

Using the results of the previous appendix we here prove the theorem that allows us to identify the low energy states as all reorderings of the $\mf h$, $\mf p$ that make 
up the ground state. 

\textbf{Lemma:}
For all $n,m\in\mathbb{Z}$
\begin{eqnarray}
  \left(\tilde{\mf{h}}_{k\pm\frac{1}{2}R}\mf{p}^n_{kR}\tilde{\mf{p}}_{kR}\right)^T & = 
  & \tilde{\mf{h}}_{k\pm\frac{1}{2}R}\mf{p}^n_{kR}\tilde{\mf{p}}_{kR}\label{eq:trans:hp^m=hp^m} \\
  \left(\tilde{\mf{h}}_{rR}\mf{h}^m_{rR}\tilde{\mf{p}}_{r\pm\frac{1}{2}R}\right)^T & = 
  & \tilde{\mf{h}}_{rR}\mf{h}^m_{rR}\tilde{\mf{p}}_{r\pm\frac{1}{2}R}\label{eq:trans:h^mp=h^mp}
\end{eqnarray}

\emph{Proof:}
Since $\mf{h}_{k\pm\frac{1}{2}R}\mf{p}_{kR}$ is a unit cell, $\mf{h}_{k\pm\frac{1}{2}R}\mf{p}^{n+1}_{kR}$ is also a unit cell, $\forall n\in\mathbb{Z}$.
For any unit cell $C_R=0\tilde{C}1$, $\tilde C^T=\tilde C$ (property \ref{symmetry} above),
then $\left(\tilde{\mf{h}}_{k\pm\frac{1}{2}R}\mf{p}^n_{kR}\tilde{\mf{p}}_{kR}\right)^T = \tilde{\mf{h}}_{k\pm\frac{1}{2}R}\mf{p}^n_{kR}\tilde{\mf{p}}_{kR}$
which is \pref{eq:trans:hp^m=hp^m}.
Equation \pref{eq:trans:h^mp=h^mp} follows analogously.
\begin{flushright}$\square$\par\end{flushright}

Using this lemma we get:

\textbf{Lemma:}
\begin{eqnarray}
  \tilde{\mf{h}}_{kL}1 & = & 1\tilde{\mf{h}}_{kR}\label{eq:h_l1--1h_r}\\
  0\tilde{\mf{p}}_{rL} & = & \tilde{\mf{p}}_{rR}0\label{eq:0p_l--p_r0}
\end{eqnarray}

\emph{Proof:}
Using the transpose of \pref{eq:p_R_decompose}  we find
\begin{eqnarray*}
\tilde{\mf{h}}_{kL}1 & = & 1\tilde{\mf{p}}_{k+\frac{1}{2}L}\mf{p}_{k+\frac{1}{2}L}^{n-1}\tilde{\mf{h}}_{k+1L}1\\
& = & 1\tilde{\mf{h}}_{k+1R}\mf{p}_{k+\frac{1}{2}R}^{n-1}\tilde{\mf{p}}_{k+\frac{1}{2}R}1\\
& = & 1\tilde{\mf{h}}_{kR} \ ,\end{eqnarray*}
where we used \eqref{eq:trans:hp^m=hp^m} (and property \ref{enu:transposition}) in the second step. The proof for \eqref{eq:0p_l--p_r0} is analogous to that of \eqref{eq:h_l1--1h_r}.
\begin{flushright}
$\square$
\par\end{flushright}
One trivial consequence of this lemma is that $\left(\tilde{\mf h}_{kL}1\right)^T = \tilde{\mf h}_{kL}1$ and $\left(0\tilde{\mf p}_{rL}\right)^T= 0\tilde{\mf p}_{rL}$.

\textbf{Lemma:}
\begin{equation}
\tilde{\mf{h}}_{kL}10\tilde{\mf{p}}_{k\pm\frac{1}{2}L}=\mf{p}_{k\pm\frac{1}{2}L}\mf{h}_{kL}\label{eq:particle_hole_switch}\end{equation}

\emph{Proof:} 
$\tilde{\mf h}_{kL}10\tilde{\mf{p}}_{k\pm\frac{1}{2}L}=1\tilde{\mf{h}}_{kR}\tilde{\mf{p}}_{k\pm\frac{1}{2}R}0
=1\tilde{\mf p}_{k\pm\frac{1}{2}L}\tilde{\mf{h}}_{kL}0=\mf{p}_{k\pm\frac{1}{2}L}\mf h_{kL}$
as a direct consequence of first  \eqref{eq:h_l1--1h_r} and \eqref{eq:0p_l--p_r0}, then \eqref{eq:trans:hp^m=hp^m}  and finally property \ref{enu:def-tilde}.
\begin{flushright}$\square$\par\end{flushright}

\textbf{Lemma:}
\begin{eqnarray}
\mf{p}_{kR} & = & \left\{ \dots\right\} 1\tilde{\mf{h}}_{k+\frac{1}{2}R}\label{eq:P_r--<..>h_or}\\
\mf{h}_{kR} & = & \tilde{\mf{p}}_{k+\frac{1}{2}R}0\left\{\dots \right\} \label{eq:H_r--p_or<..>}\end{eqnarray}
where $\left\{\dots \right\}$ stands for something undetermined.

\emph{Proof:}
Using the decomposition \pref{eq:p_R_decompose} of $\mf h$ and $\mf p$ and  \eqref{eq:trans:h^mp=h^mp}, we get 
\begin{eqnarray*}
\mf{p}_{kR} & = & \mf{h}_{k+\frac{1}{2}R}^n\mf{p}_{k+1R}=0\tilde{\mf{h}}_{k+\frac{1}{2}R}\mf{h}_{k+\frac{1}{2}R}^{n-1}\tilde{\mf{p}}_{k+1R}1\\
& = & 0\tilde{\mf{p}}_{k+1L}\mf{h}_{k+\frac{1}{2}L}^{n-1}\tilde{\mf{h}}_{k+\frac{1}{2}L}1=\left\{\dots \right\} \tilde{\mf{h}}_{k+\frac{1}{2}L}1\\
 & = & \left\{\dots \right\} 1\tilde{\mf{h}}_{k+\frac{1}{2}R} \ .
 \end{eqnarray*}
The proof of \eqref{eq:H_r--p_or<..>} is analogous.

\begin{flushright}$\square$\par\end{flushright}

\textbf{Theorem:}\label{perm_theorem}

Assume a representation $C_{\left(\delta\right)}$ with the corresponding
particle-hole decomposition $C_{\left(\delta\right)}=\mf{hp}$
exists. Assume that a state, $S$, can be expressed as a sequence of
only $\mf{p}$ and $\mf{h}$, $S=\eta_{1}\dots\eta_{N}$,
where $\eta_{j}\in\left\{ \mf{p},\mf{h}\right\} $. Then
any permutation $S^{\prime}=\sigma\left(S\right)=\eta_{1}^{\prime}\dots\eta_{N}^{\prime}$,
where $\eta_{j}\in\left\{ \mf{p},\mf{h}\right\} $, differs from
$S$ in energy with terms of order $q$ or higher, {\it  ie}
\begin{equation*}
\Delta E=E\left(S^{\prime}\right)-E\left(S\right)=\mathcal{O}\left(q \right) \ . 
\end{equation*}
Any  other state $R$, {\it ie} a state that cannot be obtained by reordering the $\mf p$ and $\mf h$ in the ground state has energy
\begin{equation*}
\Delta E=E\left(R\right)-E\left(S\right)=\mathcal{O}\left(k \right) \  k< q  \ . 
\end{equation*}

\emph{Proof:}
We start by proving, by induction, the first part of the theorem.
Any permutation $\sigma$ can be written as a series of (pairwise) transpositions $\tau_{k}$.
If all transpositions fulfill the requirement above then so does any permutation.

We prove this using the $C_{R}$-representation, but the result is valid for all representations since they are equivalent.
(Equivalent in the sense that the same sequence of $\mf p$ and $\mf h$, in different representations,
gives (under periodic boundary conditions) the same sequence of 0 and 1, upp to a translation of $<q$ sites).
In this representation $\mf{ph}\rightarrow\mf{hp}$ is affected using $\mf p_R\mf{h}_{R}=\tilde{\mf{p}}_{R}10\tilde{\mf{h}}_{R}\rightarrow\tilde{\mf{p}}_{R}01\tilde{\mf{h}}_{R}=\mf{h}_{R}\mf p_R$
as is seen from (the transpose of) \eqref{eq:particle_hole_switch}.

We have four cases that need to be treated separately.
$\mf{h}\overleftrightarrow{\mf{ph}}\mf{p}$, $\mf{p}\overleftrightarrow{\mf{ph}}\mf{h}$,
$\mf{h}\overleftrightarrow{\mf{ph}}\mf{h}$ and $\mf{p}\overleftrightarrow{\mf{ph}}\mf{p}$.
The transpositions of these four  $\mf{p}\overleftrightarrow{\mf{h}\mf{p}}\mf{h}$, etc,  are treated in analogy with the first four.

We show only the proof for $l_{\mf p}<l_{\mf h}\Rightarrow\mf h_R=\mf h_{1R}\mf p_R^n$;
the case $l_{\mf p}>l_{\mf h}\Rightarrow\mf p_R=\mf h_R^m\mf p_{1R}$ is analogous and 
$l_{\mf p}=l_{\mf h}\Rightarrow\mf p=1,\mf h=0$ is trivial.
Here $\mf{h}_{R}$ and $\mf p_R$ are shorthands for $\mf{h}_{0R}$ and $\mf{p}_{\frac{1}{2}R}$ respectively.
Also $\mf h'_R=\mf h_{1R}\mf p^{n-1}_R$ and $\mf p'_R=\mf h^{m-1}_{1R}\mf p_{\frac{3}{2}R}$, which are the minimal expansions of $\mf h _R$ and $\mf p_R$, do appear.

Consider the first case:
\begin{eqnarray*}
\left\{ \dots \right\}\mf{h}_{R}\overleftrightarrow{\mf p_R\mf{h}_{R}}\mf p_R & = & \left\{ \dots 1 \right\} 0\tilde{\mf{h}}_{R}\tilde{\mf{p}}_{R}\overrightarrow{10}
\tilde{\mf{h}}_{R}\tilde{\mf{p}}_{R}1\\
 & = & \left\{ \dots1\right\} 0\left\langle q-2\right\rangle \overrightarrow{10}
\left\langle q-2\right\rangle^T 1 \ ,
 \end{eqnarray*}
where $\left\langle q-2\right\rangle =\tilde{\mf h}_R \tilde{\mf p}_R=\left\langle q-2\right\rangle^T$ according to \pref{eq:trans:hp^m=hp^m}.
$\left\{\dots \right\}$ denotes an arbitrary string of $\eta$'s -- in the $R$-representation it necessarily ends with a 1;
this string is included since it gives the leading contribution to the energy.
$\left\langle n\right\rangle$ denotes a string that is $n$ sites long such that 
$\left\langle n\right\rangle x\left\langle n \right\rangle^T$ is symmetric around $x$.
Because of this reflection symmetry, $\left\langle n \right\rangle 10\left\langle n \right\rangle^T$ and
$\left\langle n \right\rangle01\left\langle n \right\rangle^T$ have the same energy.
It then follows that the leading contribution to the energy is given by the interaction with the next electron on both sides and is $V_q^{\prime\prime}$,
thus $\Delta E=V_{q}^{\prime\prime}+\mathcal{O}\left(q+1\right)$.

The second case:
\begin{eqnarray*}
  \mf p_R\overleftrightarrow{\mf p_R\mf h_R}\mf{h}_{R} 
  & = & \left\{\mf p_R^n\right\}\mf p_R\tilde{\mf p}_R\overrightarrow{10}
  \tilde{\mf h}'_R\tilde{\mf p}_R10\tilde{\mf h}_R\\
  & = & \left\{\mf p_R^n\right\}\mf p_R\tilde{\mf p}_{R}\overrightarrow{10}
  \tilde{\mf p}_L\tilde{\mf h}'_L10\tilde{\mf h}'_R\tilde{\mf p}_R1\\
  & = & \left\{\mf p_R\mf p_R\right\}\left<l_{\mf p^n}-1\right>\overrightarrow{10}
  \left<l_{\mf p^n}-1\right>^T\\
  & & \hphantom{\mf{ppppppppppp}}
  \tilde{\mf h}_{1L}10\tilde{\mf p}_L\tilde{\mf h}'_L1\\
  & = & \left\{\mf p_R\mf p_R\right\}\left<l_{\mf p^n}-1\right>\overrightarrow{10}
  \left<l_{\mf p^n}-1\right>^T \\
  & & \hphantom{\mf{ppppppppppp}}
  \mf p_L\mf h_{1L}\tilde{\mf h}'_L1\\
  & = & \left\{\mf p_R\right\}\left<l_{\mf p^{n+1}}-1\right>\overrightarrow{10}
  \left<l_{\mf p^{n+1}}-1\right>^T\mf h_{1L}\tilde{\mf h}'_L1\\
  & = & \left\{\ldots1\tilde{\mf h}_{1R}\right\}\left<l_{\mf p^{n+1}}-1\right>\overrightarrow{10} \\
  & & \hphantom{\mf{ppppppppppp}}
  \left<l_{\mf p^{n+1}}-1\right>^T\tilde{\mf h}_{1L}01\tilde{\mf h}'_R\\
  & = & \left\{\ldots1\right\} \left<q-2\right>\overrightarrow{10}
  \left<q-2\right>^T01\tilde{\mf h}'_R \ ,
\end{eqnarray*}
where, in the second to last step we used $\mf p_R=0\tilde{\mf h}_{1R}\tilde{\mf p}'_R1=0\tilde{\mf p}'_L\tilde{\mf h}_{1L}1=0\tilde{\mf p}'_L1\tilde{\mf h}_{1R}$.
We find that the nearest electrons are $q-1$ sites to the left and $q+1$ sites to the right, thus $\Delta E=-V_{q}^{\prime\prime}+\mathcal{O}\left(q+1\right)$.

The third case, using \pref{eq:trans:hp^m=hp^m},
\begin{eqnarray*}
  \mf p_R\overleftrightarrow{\mf p_R\mf h_R}\mf p_R & = 
  & \left\{\mf p_R^n\right\}\mf p_R\tilde{\mf p}_R\overrightarrow{10}
  \tilde{\mf h}'_R\mf p_R\tilde{\mf p}_R1\\
  & = & \left\{\mf p_R^n\right\} \mf p_R\tilde{\mf p}_{R}\overrightarrow{10}
  \tilde{\mf p}_L\mf p_L\tilde{\mf h}'_L1\\
  & = & \left\{\mf p_R\right\} \left<l_{\mf p^{n+1}}-1\right>\overrightarrow{10} \\
  & & \hphantom{\mf{ppppppppppp}}
  \left<l_{\mf p^{n+1}}-1\right>^T\tilde{\mf{h}}_{1L}1\\
  & = & \left\{\ldots1\tilde{\mf{h}}_{1R}\right\} \left<l_{\mf p^{n+1}}-1\right>\overrightarrow{10} \\
  & & \hphantom{\mf{ppppppppppp}}
  \left<l_{\mf p^{n+1}}-1\right>^T\tilde{\mf{h}}_{1L}1\\
  & = & \left\{\ldots1\right\}\left<q-2\right>\overrightarrow{10}\left<q-2\right>^T1  \ .
\end{eqnarray*}
Here the nearest pair of electrons that contributes is more than $q-1$ sites to the left and more than $q$ sites to the  right, thus $\Delta E=\mathcal{O}\left(q+1\right)$.

Finally, the fourth case
\begin{eqnarray*}
  \mf{h}_{R}\overleftrightarrow{\mf p_R\mf h_R}\mf h_R 
  & = & \left\{\mf p_R\right\}\mf h_R\tilde{\mf p}_R\overrightarrow{10}
  \tilde{\mf h}'_R\tilde{\mf p}_R10\tilde{\mf h}'_R\tilde{\mf p}_R1\\
  & = & \left\{\mf p_R\right\}\mf h_R\tilde{\mf p}_R\overrightarrow{10}
  \tilde{\mf p}_L\tilde{\mf h}'_L10\tilde{\mf p}_L\tilde{\mf h}'_L1\\
  & = & \left\{\mf p_R\right\}\mf h_R\left<l_{\mf p}-1\right>\overrightarrow{10} \\
  & & \hphantom{\mf{ppppppppppp}}
  \left<l_{\mf p}-1\right>^T\tilde{\mf h}'_L10\tilde{\mf p}_L\tilde{\mf h}'_L1\\
  & = & \left\{\mf p_R\right\}\mf h_R\left<l_{\mf p}-1\right>\overrightarrow{10} \\
  & & \hphantom{\mf{ppppppppppp}}
  \left<l_{\mf p}-1\right>^T\mf p_L\mf h'_L\tilde{\mf h}'_L1\\
  & = & \left\{\tilde{\mf p}'_R1\right\}\left<q-1\right> \overrightarrow{10}\left<q-1\right>^T\tilde{\mf h}'_L1 \ .
\end{eqnarray*}
The nearest electrons are here $q$ sites to the left and more than $q$ sites to the right, thus $\Delta E=\mathcal{O}\left(q+1\right)$.

Since all transpositions cost energy $\Delta E=c_{q}V_{q}^{\prime\prime}+\mathcal{O}\left(q+1\right)$,
the energy cost for any permutation must also be of this form.
\begin{flushright}$\square$\par\end{flushright}

This finishes the proof of the first part of the theorem. We now turn to the second part which states that all states that cannot be obtained by reordering the $\mf p$ and $\mf h$ that 
make up the ground state differ in energy from the ground state by $\mathcal{O}\left(k\right), \, k<q$.

Assume that we have a state $S_0$ that can be described by $\mf{p,h}$.
We say that $\mf{p,h}$ covers $S_0$ if and only if we can describe $S_0$ using $\mf{p,h}$.
$S_0$ is our reference state; it could be the ground state.

Let $S$ be  a state that is not covered by $\mf{p,h}$. Assuming $l_{\mf p} < l_{\mf h}$, we  expand $\mf{ph}=\mf{pph'}$.
Note that $\mf{p,h'}$ trivially cover $S_0$ since $\mf{p,h}$ do.
If $\mf{p,h'}$ cover $S$ then
the sequence $\mf{h'h'}$ must occur somewhere in $S$ otherwise $\mf{h,p}$ would also cover $S$.
This means that when $S_0 \rightarrow S$ there has been a move of the type 
\begin{equation*}
\mf{h'p}^{n+1}\mf{h'p}\rightarrow\mf p^n\mf{h'ph'p}\rightarrow\mf p^n\mf{h'h'p} \ ,
\end{equation*}
where the last step costs $V''_{l_{\mf h}}$ as seen in the proof the first part of the  theorem. Since $l_{\mf h}<q$, this proves the theorem. 
If $\mf{p,h'}$ do not cover $S$, then expand the longer of $\mf h'$ and $\mf p$ and repeat the argument. Eventually a particle and a hole is 
found that covers $S$ and the theorem follows.
  
Defining the low energy states as those whose energy are within $\mathcal{O}\left(q\right)$
from the ground state energy we find that the low energy states
are precisely the states $S=\eta_{1}\dots\eta_{N}$ where $\eta_{j}\in\left\{ \mf p,\mf{h}\right\} $.

\section{Exclusion statistics}\label{app:exclusion}

Assuming the low-energy sector identified in App. \ref{app:ph-proofs}  it is straightforward to derive the exclusion statistics in the TT-limit,
thus generalizing the result in \cite{we} to general $\nu=p/q$. 

Let the ground state contain $N_c$ unit cells $C$:  $\mf{hphphp}\dots\mf{hphphp}$.
Let $N_{\mf p}$, $N_{\mf h}$ denote the number of $\mf p$, $\mf h$ in a general state;
for the ground state $N_{\mf p}=N_{\mf h}=N_c$.
A quasielectron $\mf p$ (quasihole $\mf h$) can then be inserted once per quasihole $\mf h$ (quasielectron $\mf p$) to produce different quasielectron (quasihole) states.
Thus the number of one-quasielectron states is equal to the number of quasiholes $N_{\mf h}$ present, $N_{\mf h}=N_c$, 
and the number of one-quasihole states is equal to the number of quasielectrons $N_{\mf p}$ present, $N_{\mf p}=N_c$.
To determine how this changes in the presence of quasiparticles,
we insert $q$ quasielectrons $\mf p$ in the ground state and remove $l_{\mf p}$ unit cells $C$ in order for the size of the system to remain unchanged.
There are now $N^\prime_{\mf p}=N_c+q-l_{\mf p}=N_c+l_{\mf h}$ quasielectrons $\mf p$ and $N^\prime_{\mf h}=N_c-l_{\mf p}$ quasiholes $\mf h$ present in the state.
A quasielectron $\mf p$ can still be inserted once per quasihole $\mf h$ to produce different one-quasielectron states.
Thus there are now $N^\prime_{\mf h}=N_c-l_{\mf p}$ one-quasielectron states and $N^\prime_{\mf p}=N_c+l_{\mf h}$ one-quasihole states.
This gives  $g_{\mf{pp}}=-\frac{\Delta N_{\mf h}}{q}=\frac{l_{\mf{p}}}{q}$ and 
$g_{\mf{hp}}=-\frac{\Delta N_{\mf p}}{q}=-\frac{l_{\mf h}}{q}$ because we actually inserted 
$q$ quasielectrons $\mf p$.
Inserting instead $q$ quasiholes in the ground state leads to $g_{\mf{h}\mf{h}}=\frac{l_{\mf{h}}}{q}$ and $g_{\mf{p}\mf{h}}=-\frac{l_{\mf{p}}}{q}$.
The lenghts of the quasiparticles are given by \pref{ln}.

We note that while the diagonal parameters  are positive as they should be in order for the number of states 
of a particular species not to increase with the addition of particles of this species, the off-diagonal parameters are always negative.
In the TT-limit  we have that $g_{--}+g_{++}=1$, $g_{+-}=-g_{++}$ and $g_{-+}=-g_{--}$, which differs from the results  $g_{--}+g_{++}=2$ and  $g_{+-}=-g_{-+}$, 
that hold away from this limit \cite{exclusion,isakov}.
This discrepancy is further discussed in Sec. \ref{Numerics}.

When calculating the statistics parameters above the size of the system was maintained. 
However, when comparing to exact diagonalisation of small systems this constraint needs to be relaxed.
We do this by first considering how the corresponding constraint can be relaxed when calculating the fractional charge of the quasiparticles.
Normally, one adds several quasiparticles and removes an integral number of unit cells so that the size of the system is unchanged. 
We can however reformulate the calculation and insert just one $\alpha$ particle.
This particle contains $n_{\alpha}$ electrons and extends over $l_{\alpha}$ sites. 
To obtain the charge of the quasiparticle we must subtract the charge of the background, which has charge $p$ per unit cell of length $q$.
Thus we find that the charge of the $\alpha$ quasiparticle is $Q_{\alpha}=n_{\alpha}-l_{\alpha}\frac{p}{q}=\frac{\alpha}{q}$,
where we have used \pref{eq:main-equation} wich $a=\alpha$.

A similar reasoning can be applied to the calculation of the exclusion statistics.
Consider the TT-ground state with $N_c$ unit cells at filling factor $\nu=p/q$.
We now insert one $\alpha$ particle ($\alpha=\mf p$ or $\alpha=\mf h$).
Let $N_\alpha$ be the number of  low-energy states thus obtained. 
(If there is one such state per unit cell, which is typically the case, then these $N_\alpha$ states are all degenerate.)
We now, in another TT-ground state of $N_c$ unit cells, insert instead a $\beta$-particle.
($\alpha$ and $\beta$ could be the same kind of particle.)
Here we will get $N_\beta$ low energy states, just like for $\alpha$ above.
We fix the position of the first inserted $\beta$-particle and insert an $\alpha$-particle.
Let $N_{\alpha \beta}$ be the number of low-energy states thus obtained. 
(These low energy states  will not be degenerate since the first insertion breaks the translational invariance.)
The difference $\Delta_{\alpha \beta} = N_{\alpha \beta}-N_\alpha$ in the number of low energy states before and after the insertion of the $\beta$ particle tells us
how much the state space of $\alpha$ changes by the insertion of a $\beta$. We obtain the statistics parameter $g_{\alpha \beta}$ from $\Delta_{\alpha \beta}$ by
compensating for the change in size of the system the insertion of the $\beta$-particle gave rise to. 
We know that a unit cell contains exactly one particle and one hole,
therefore the number of particles and holes per site is $1/q$.
We added a $\beta$-particle containing $l_\beta$ sites and by that very reason the number of low energy states for the particle and hole increased by ${l_\beta}/{q}$ respectively.
Deducting this modification from the number of low-energy states we find the exclusion statistics   
$g_{\alpha \beta}=-(\Delta_{\alpha \beta}-\frac{l_\beta}{q})=\frac{l_\beta}{q}-\Delta_{\alpha \beta}$.

Applying the above procedure to the TT-limit gives just another way of deriving the  $g_{\alpha \beta}$ given above. Here we apply it to obtain the exclusion statistics 
for small systems analyzed using exact diagonalization.

We start with $\nu=1/3$.
Fig. \ref{fig:5_15_h_hh} shows the insertion of one (5/16) and two (5/17) holes at $\nu=1/3$ with $N_c=5$ unit cells.
The TT-ground state at $5/15$ is $(\mf{hp})_5$.
Adding one hole one obtains 5/16, where the lowest energy state in Fig. \ref{fig:5_15_h_hh} is $\mf h_2\mf p(\mf{hp})_4$ in the TT-limit.
There are obviously 16 such states related by translation, but choosing one particular ground state initially,
the additional $\mf h$ can be inserted in five different places; thus, $N_{\mf h}=5$.
At 5/17, the three lowest energy states in Fig. \ref{fig:5_15_h_hh} are found to be $\mf h(\mf{hp})_2\mf h(\mf{hp})_3$,
$\mf h(\mf {hp} ) \mf h (\mf{hp})_4$ and $\mf h_2(\mf{hp})_5$ (each of them 17-fold degenerate) in the TT-limit, in order of increasing energy.
To determine $N_{\mf{hh}}$ we must keep the first hole fixed and count the number of states for the second hole $\mf h^{\prime}$---the three low-energy states (and translations of them)
then give the states:
$\mf h\mf h^\prime(\mf{hp})_5$, $\mf{h(hp)h^\prime(hp)}_4$, $\mf{h(hp)}_2\mf{h^\prime(hp)}_3$,
$\mf{h(hp)}_3\mf{h^\prime(hp)}_2$, and $\mf{h(hp)}_4\mf{h^\prime (hp)}$; thus $N_{\mf{hh}}=5$.
($\mf h$ and $\mf h^\prime$ are of course identical so the order of them does not matter.)
Of the five states, there are two pairs that are related by translation and hence degenerate.
Since the numbers of low-energy quasihole states, $N_{\mf h}$ and $N_{\mf{hh}}$, 
are equal, we conclude that $\delta_{\mf{hh}}=0$. 
The TT-limit is particle-hole symmetric, thus the analysis for insertion of particles, {\cf} Fig. \ref{fig:5_15_p_pp}, is obtained by simply interchanging $\mf h$ and $\mf p$, 
consequently $\delta_{\mf{pp}}=0$. 
Since $l_{\mf p}=2$ and $l_{\mf h}=1$ we get $g_{\frac{1}{3}\mf{pp}}=\frac{2}{3}$
and $g_{\frac{1}{3}\mf{hh}}=\frac{1}{3}$.

Fig. \ref{fig:4_10_h_hh} shows the insertion of one (5/13) and two (6/16) holes at $\nu=2/5$ with $N_c=2$ unit cells. 
The TT-ground state at $\nu=4/10$ is $\mf{(hp)}_2$. Adding one hole one obtains 5/13, where the lowest state is $\mf h_2\mf{php}$.
There are obviously 13 such states related by translation, but choosing one particular ground state initially, the additional $\mf h$ can be inserted in two different places; thus, $N_{\mf h}=2$.
At $6/16$, the two lowest energy states are found to be $(\mf h_2\mf p)_2$  and $\mf h_2(\mf{hp})_2$, in the TT-limit, in order of increasing energy.
The first of these is 8-fold degenerate whereas the second is 16-fold. 
(The additional degeneracy disappears when $L_1$ increases and linear combinations of states with different positions of the holes are formed.)
To determine $N_{\mf{hh}}$ we keep the first hole fixed and count the number of states for the second hole $\mf h^{\prime}$---these states are:
$\mf{hh^\prime (hp)}_2$ and $\mf{h(hp)h^\prime (hp)}$; thus $N_{\mf{hh}}=2$. 
Since the numbers of low-energy hole states, $N_{\mf h}$ and $N_{\mf{hh}}$, are equal, we conclude that $\delta_{\mf{hh}}=0$. 
The TT-limit is particle-hole symmetric, thus the analysis for insertion of particles,
{\cf} Fig. \ref{fig:4_10_p_pp}, is obtained by simply interchanging $\mf h$ and $\mf p$, 
consequently $\delta_{\mf{pp}}=0$. 
Since $l_{\mf p}=2$ and $l_{\mf h}=3$ we get $g_{\frac{2}{5}\mf{pp}}=\frac{2}{5}$ and $g_{\frac{2}{5}\mf{hh}}=\frac{3}{5}$.

The off-diagonal elements of the statistics parameter are obtained from Fig. \ref{fig:4_10_ph_5_15_ph},
which shows the spectra for $\nu=1/3$ with six unit cells and $\nu=2/5$ with three cells.
At 6/18 the lowest energy states are identified as $(\mf{hp})_6$, $\mf{h(hp)p(hp)}_4$,
$\mf{h(hp)}_2\mf{p(hp)}_3$, $\mf{h(hp)}_3\mf{p(hp)}_2$, $\mf{h(hp)}_4\mf{p(hp)}$
and $\mf{h(hp)}_5\mf p$ in the TT-limit  (in order of increasing energy).
(Four of these states are pairwise degenerate.)
The first of these is of course the 1/3 ground state, whereas the others are particle-hole excitations thereof.
All these states can be obtained by inserting a hole and a particle in the 5/15 ground state $(\mf{hp})_5$.
Adding one hole to $(\mf{hp})_5$ one obtains the five low energy states $\mf{h(hp)}_5$ according to the analysis above, thus $N_{\mf h}=5$.
Adding a particle, keeping the added hole fixed, then gives precisely the six low-energy states found at 6/18.
Thus there are $N_{\mf{hp}}=6$ particle states. 
Thus, when we move from $\frac{5}{15}+\mf h$ to $\frac{5}{15}+\mf h+\mf p$ the number of states
increases by one from 5 to 6, $\delta_{\mf{ph}}=1$,
(the process $\frac{5}{15}+\mf{p}$ to $\frac{5}{15}+\mf p+\mf h$ of course gives the same counting).
This gives the off-diagonal elements $g_{\frac{1}{3}\mf{hp}}=\frac{2}{3}-1=-\frac{1}{3}$
and $g_{\frac{1}{3}\mf{ph}}=\frac{1}{3}-1=-\frac{2}{3}$.

For $\nu=\frac{2}{5}$ at 6/15, {\it cf} Fig.  \ref{fig:4_10_ph_5_15_ph}, we identify the low-energy states as $(\mf{hp})_3$, $\mf{h(hp)p(hp)}$, and $\mf{h(hp)_2p}$ 
 in the TT-limit (the last two being degenerate). By a similar analysis as for 1/3 this gives again $\delta_{\mf{hp}}=1$,
and hence $g_{\frac{2}{5}\mf{hp}}=\frac{2}{5}-1=-\frac{3}{5}$
and $g_{\frac{2}{5}\mf{ph}}=\frac{3}{5}-1=-\frac{2}{5}$.


\begin{thebibliography} {999}

{\footnotesize

\bibitem{bk1} E.J. Bergholtz, and A. Karlhede, Phys. Rev. Lett. {\bf 94}, 26802 (2005).

\bibitem{Lee05} A. Seidel, H. Fu, D.-H. Lee, J.M. Leinaas, and J. Moore, Phys. Rev. Lett. {\bf 95}, 266405 (2005).


\bibitem{bk2} E.J. Bergholtz, and A. Karlhede, J. Stat. Mech. (2006) L04001.

\bibitem{emil} E.J. Bergholtz, and A. Karlhede, Phys. Rev. B {\bf 77}, 155308 (2008).

\bibitem{Jansen} S. Jansen, E.H. Lieb,  and R. Seiler, Commun. Math. Phys. 285, 503 (2009); phys. stat. sol. (b) 245, No. 2, 439 (2008).

\bibitem{wehierarchy} E.J. Bergholtz, T.H. Hansson, M. Hermanns and A. Karlhede, Phys. Rev. Lett. {\bf 99}, 256803 (2007). 

\bibitem{exclusion} F.D.M. Haldane, Phys. Rev. Lett. {\bf 67}, 937 (1991).

\bibitem{we} T.H. Hansson, and A. Karlhede, arXiv:0907.0672 (2009).

\bibitem{canright} M.D. Johnson, and G.S. Canright, Phys. Rev. B {\bf 49}, 2947 (1994).

\bibitem{canright94} G.S. Canright, and M.D. Johnson, J. Phys. A {\bf 27} 3579 (1994).

\bibitem{wu94} Y.S. Wu, Phys. Rev. Lett. {\bf 73}, 922 (1994).

\bibitem{su}W.-P. Su, Y.-S Wu, and J. Yang, Phys. Rev. Lett. {\bf 77}, 3423 (1996). 

\bibitem{isakov} S.B. Isakov, G.S. Canright, and M.D. Johnson,  Phys. Rev. B {\bf 55}, 6727 (1997).

\bibitem{sitko03} P. Sitko, Phys. Rev. B {\bf 68}, 153309 (2003).

\bibitem{hermanns10} M. Hermanns, A. Chandran, N. Regnault, and B. A. Bernevig, arXiv:1009.4199v1 (2010). 

\bibitem{haldane08} H. Li, and F.D.M. Haldane, Phys. Rev. Lett. {\bf 101}, 010504 (2008).

\bibitem{bergholtz10} A.M. L\"auchli, E.J. Bergholtz, J. Suorsa, and M. Haque, Phys. Rev. Lett. {\bf 104}, 156404 (2010). 

\bibitem{SeidelLee} A. Seidel, and D.-H. Lee, Phys. Rev. B {\bf 76}, 155101 (2007). 

\bibitem{Seidel08} A. Seidel, Phys. Rev. Lett. {\bf 101}, 196802 (2008).




}


\end{thebibliography}
\end{document}